\documentclass[a4paper,11pt]{article}
\pdfoutput=1
\usepackage{jheppub}
\usepackage{amsmath}
\usepackage{hyperref}
\usepackage{soul}
\usepackage{cancel}
\usepackage[normalem]{ulem}
\allowdisplaybreaks
\def\bea#1\eea{\begin{align}#1\end{align}}
\def \be {\begin{equation}}
\def \ee {\end{equation}}
\newcommand{\nnu}{\nonumber\\}

\newcommand{\bef}{\begin{figure}[hbt]\centering}
\newcommand{\eef}{\end{figure}}
\newcommand{\sla}[1]{{#1}\!\!\!\slash}

\newcommand\as{\alpha_s}
\newcommand{\f}{\frac}

\newcommand{\cc}[1]{\mathcal{#1}}
\preprint{YITP-SB-20-13}

\title{Power expansion for heavy quarkonium production at next-to-leading order in $\rm e^+e^-$ annihilation}

\author{Kyle Lee}
\author{and George Sterman}

\affiliation{C.N. Yang Institute for Theoretical Physics and Department of Physics and Astronomy, \\Stony Brook University, \\Stony Brook, NY 11794, U.S.A.}
  
\emailAdd{kunsu.lee@stonybrook.edu}
\emailAdd{george.sterman@stonybrook.edu}

\abstract{We study heavy quarkonium production associated with gluons in $\rm e^+e^-$ annihilation as an illustration of the perturbative QCD (pQCD) factorization approach, which incorporates the first nonleading power in the energy of the produced heavy quark pair.    We show how the renormalization of the four-quark operators  that define the heavy quark pair fragmentation functions using dimensional regularization induces ``evanescent"  operators that are absent in four dimensions.  We derive closed forms for short-distance coefficients for  quark pair production to next-to-leading order ($\alpha_s^2$)
 in the relevant color singlet and octet channels.    Using non-relativistic QCD (NRQCD) to calculate the heavy quark pair fragmentation functions up to $v^4$ in the velocity expansion, we derive analytical results for the differential energy fraction distribution of the heavy quarkonium.   Calculations for ${}^3S_1^{[1]}$ and ${}^1S_0^{[8]}$ channels agree with analogous NRQCD analytical results available in the literature, while several color-octet calculations of energy fraction distributions are new. We show that the remaining corrections due to the heavy quark mass fall off rapidly in the energy of the produced state. To explore the importance of evolution at energies much larger than the mass of the heavy quark, we solve the renormalization group equation perturbatively to two-loop order for the ${}^3S_1^{[1]}$ case.}

\begin{document}
\maketitle

\section{Introduction}

Heavy quarkonium production is a subject of continuing interest  \cite{Bodwin:1994jh,Brambilla:2010cs,Brambilla:2004wf}.  The production of a heavy quarkonium state always involves an intrinsic hard scale, the heavy quark mass, $m_Q$. In the presence of an  even larger hard scale, $E_H$, such as transverse momentum or particle energy, large logarithms $\ln (E_H^2/(2m_Q)^2)$ can appear, which are fully perturbative but require resummation. The perturbative QCD (pQCD) factorization procedure developed in refs.\ \cite{Kang:2014tta, Kang:2014pya} is an organization of high-energy quarkonium production into subprocesses of different characteristic regions in momentum and coordinate space to make this resummation possible. Closely-related results have been derived from an effective theory viewpoint in refs.\ \cite{Fleming:2012wy,Fleming:2013qu}. Recently, the formalism also has been applied to light meson production in deep-inelastic scattering \cite{Liu:2019srj}.

For quarkonium production at large transverse momentum, this pQCD factorization approach  provides a unified framework for leading-power (LP) and next-to-leading power (NLP) behavior in momentum transfer for production cross sections. Suppressing convolutions associated with the initial state, such factorized cross sections for the production of heavy quarkonium $H$ can be represented as 
\bea
\label{eq:QCDfact}
\sigma^{H}_{\text{pQCD}} = \underbrace{\sum_{i=q,\bar{q},g}\hat{\sigma}_{i} \otimes_z D_{i\to H}}_\text{LP} + \underbrace{\sum_\kappa \hat{\sigma}_{Q\bar{Q}(\kappa)} \otimes_{z;u,v} {\cal{D}}_{Q\bar{Q}(\kappa)\to H}}_\text{NLP}\,.
\eea
The convolutions in this expression are in hadronic momentum fraction $z$, and variables, $u$ and $v$, which represent the fraction of the hadron's total momentum carried by the heavy quark in the amplitude and in its complex conjugate, respectively\footnote{The variable $v$ here should not be confused with the pair relative velocity in NRQCD.
 }.
 The first term represents the LP production of a single parton $i=q,\bar q,g$ at short distances.  The second term describes the NLP production of a heavy quark pair in a specific spin and color state $\kappa$ at short distances.  
The full expression for the factorized cross section in eq.\ (\ref{eq:QCDfact}) with explicit convolutions is \cite{Kang:2014tta}
\bea
E_P\frac{d\sigma_{A+B\to H+X}}{d^3P}(P)
&=
\sum_{i=q,\bar{q},g} \int \frac{dz}{z^2} \, D_{i\to H}(z;m_Q,\mu) \, 
E_c\frac{d\hat{\sigma}_{A+B\to i(p_c)+X}}{d^3p_c}\left(p_c=\frac{1}{z}\, p,\mu \right) 
\nonumber\\
\ \nonumber\\
&\   \hspace{-10mm} +\ 
\sum_{\kappa} \int \frac{dz}{z^2} \, du\, dv\
{\cal D}_{[Q\bar{Q}(\kappa)]\to H}(z,u,v;m_Q,\mu)
\nonumber\\
&\  \hspace{-5mm}
\times \
E_c\frac{d\hat{\sigma}_{A+B\to [Q\bar{Q}(\kappa)](p_c)+X}}{d^3p_c}
\left(
p_Q=\frac{u}{z}\, p,p_{\bar{Q}}=\frac{\bar{u}}{z}\, p,
p'_Q=\frac{v}{z}\, p,p'_{\bar{Q}}=\frac{\bar{v}}{z}\, p,\mu\right)\, ,
\nonumber\\
\label{eq:pqcd_fac}
\eea
where $\mu$ is a factorization scale, and 
where here and below $\bar u\equiv 1-u$ and $\bar v\equiv 1-v$.  Throughout the paper, we work in a frame in which the heavy quarkonium momentum, $P^\mu$, is directed along the positive $z$-axis, and
denote by (lower case) $p^\mu$, a lightlike momentum with the same plus component, 
\bea
\label{eq:momenta-defs}
P^\mu\ =&\ \left( p^+,\, \frac{m_H^2}{2p^+},\, {\bf 0}_\perp \right )\, ,
\nonumber\\[2mm]
p^\mu\ =&\ \left  (p^+,\, 0,\, {\bf 0}_\perp \right ) = p^+ \bar{n}^\mu\,,
\eea
where $\bar{n}^\mu$ is the light-cone vector  $\bar{n}^\mu =  \left  (1,\, 0,\, {\bf 0}_\perp \right )$. The $p^+$ component can be projected out by the oppositely directed light-cone vector $n^\mu = \left(0,\, 1,\, {\bf 0}_\perp \right )$, as $p^+= p\cdot n$. As indicated in eq.\ (\ref{eq:pqcd_fac}), the short-distance coefficients $d\hat \sigma$ depend only on the lightlike momenta $p_Q$ and $p_{\bar{Q}}$, related to $p^\mu$ by $p = z(p_Q+p_{\bar Q})=z p_c$. That is, the heavy quark is treated as massless in the short-distance coefficients. A proof of this factorization at NLP was provided in ref.\ \cite{Kang:2014tta}. It should be noted that, rather than the full set of NLP non-perturbative functions, the formalism includes only the heavy quark pair fragmentation functions. This is justified under the reasonable assumption that they give the  dominant contributions at NLP when a heavy quarkonium is produced.    As is characteristic of factorized power corrections \cite{{Politzer:1980me,Mueller:1981sg,Jaffe:1982pm,Ellis:1982wd,Ellis:1982cd,Qiu:1988dn,Qiu:1990xxa,Qiu:1990xy}}, the fragmentation functions in this expression involve not only the momentum fraction carried by the pair, but also the relative momentum of the constituents of the pair, the heavy quark and antiquark. The short-distance functions, therefore, produce non-diagonal  products of partonic states, in which the heavy quark pair has the same total momentum, $p_c$ in eq.\ (\ref{eq:pqcd_fac}), shared differently by the heavy quark and antiquark.   The fragmentation function mediates between these products of states, non-diagonal in $u$ and $v$, and the full set of states that include the heavy quarkonium, of momentum $P$.

Operator definitions for the fragmentation functions can be found in ref.\ \cite{Kang:2014tta}, given as Fourier transforms of matrix elements of four-quark operators, here suppressing gauge links,
\bea
&\  {\cal D}_{[Q\bar{Q}(\kappa)]\to H}(z,u,v;m_Q,\mu)=
 \nonumber\\ &\  \hskip 0.45in
 \int \frac{p^+ dy^-}{2\pi}\, {\rm e}^{-i(p^+/z)y^-}
\int \frac{p^+ dy_1^-}{2\pi}\, {\rm e}^{i(p^+/z)(1-v)y_1^-}
\int \frac{p^+ dy_2^-}{2\pi}\, {\rm e}^{-i(p^+/z)(1-u)y_2^-}
\nonumber \\
&\  \hskip 0.5in
\times 
{\cal P}^{(s)}_{ij,lk}(p)\, {\cal C}^{[I]}_{ab,cd}\ \sum_X
\langle 0|\overline{\psi}_{c,l}(y_1^-)\psi_{d,k}(0)|H(P)X\rangle
\langle H(P)X| \overline{\psi}_{a,i}(y^-)\psi_{b,j}(y^- + y_2^-)|0\rangle \, ,
\nonumber\\
\label{eq:calD-def}
\eea
where ${\cal C}^{(I)}$ and ${\cal P}^{(s)}$ are members of a set of projections in color and Dirac spin space, respectively. Explicit forms of projections will be given in section \ref{sec:fact-summary}.

We will apply this formalism to heavy quarkonium production in association with gluons for $\rm e^+e^-$ annihilation. For these processes, we provide the first closed expressions for the NLP short-distance coefficients at next-to-leading order (NLO).  We will encounter a number of features that will recur in all the processes that can be factorized in this manner, and will thus play a role in any program to develop a phenomenology of heavy quark production that includes NLP mechanisms.  In addition,  once $\rm e^+e^-$ annihilation short-distance coefficients are calculated, the asymptotic behaviors of exact calculations in NRQCD at $\mathcal{O}(\alpha^2\alpha_s^2)$ allow us to check these short-distance coefficients explicitly, using the formalism for NRQCD fragmentation functions developed in refs.\ \cite{Ma:2013yla,Ma:2014eja}.  Previously, the two formalisms were compared numerically for hadronic scattering in ref.\ \cite{Ma:2014svb}, with good results, but there is nothing like an analytical comparison.

The calculation of short-distance coefficients depends, of course, on the scheme used to define the matrix elements in eq.\ (\ref{eq:calD-def}), which involve four quark fields relatively on the light cone.   As in the case of single-parton fragmentation functions, such operator configurations require renormalization as ``cut vertices" \cite{Mueller:1978xu}, as in ref.\ \cite{Kang:2014tta}, and thus depend on the renormalization scheme. In this paper, we will use a modified minimal subtraction ($\overline{\text{MS}}$) approach in dimensional regularization.   The presence of the four-quark operator then makes it necessary to include the effects of operators that are absent in a purely four-dimensional calculation.  This phenomenon is of great importance in the treatment of amplitudes mediated by effective four-quark operators \cite{Buras:1998raa,Herrlich:1994kh}. In our case, we will encounter ``evanescent'' fragmentation functions, associated with projections ${\cal P}^{(s)}$ that disappear in four dimensions. We will see that the comparison to exact NRQCD calculations, even at NLO, is ambiguous without a clear understanding of how these new functions enter into the calculation of factorized cross sections.

In this work, we study the energy fraction, $x_H= 2E_H/Q$, distributions of heavy quarkonium $H$ produced in gluon-associated processes, which only contributes at NLP up to $\mathcal{O}(\alpha^2\alpha_s^2)$. 
Note that the Belle experiment is able to separate heavy quarkonium production associated with and without $c\bar{c}$, $\sigma(e^+e^- \to J/\psi + c\bar{c})$ and $\sigma(e^+e^- \to J/\psi + X_{c\bar{c}})$ \cite{Pakhlov:2009nj}, and that the contributions without $c\bar{c}$ are mainly gluon-associated production at Belle's energy within the NRQCD formalism \cite{Abe:2001za,Aubert:2001pd,Ma:2008gq,Zhang:2009ym,Gong:2009kp,Yuan:1996ep}. The LO color octet cross section begins at order $\mathcal{O}(\alpha^2\alpha_s)$, where it would appear with an end-point enhancement \cite{Braaten:1995ez} at $x_H \sim 1$. This, however, has not been observed in the data \cite{Abe:2001za,Aubert:2001pd,Ma:2008gq,Zhang:2009ym,Gong:2009kp,Yuan:1996ep}. We focus our study in the region $x_H$ away from $1$ and thus only calculate the real contributions of the NLO short-distance coefficients. For NRQCD calculations of the gluon-associated processes, there exist NLO numerical calculations for both color-singlet channels and color-octet channels \cite{Ma:2008gq,Gong:2009kp,Zhang:2009ym}.  Closed expressions for the $x_H$ differential distributions are also available for  ${}^3S_1^{[1]}$ \cite{Cho:1996cg,Keung:1980ev,Yuan:1996ep} and ${}^1S_0^{[8]}$ \cite{Sun:2018yam} to order $\mathcal{O}(\alpha^2\alpha_s^2)$.   We will use these results to make explicit comparison to the results of pQCD factorization for these cases and see that the fixed order expression of pQCD factorization can approach the high-energy limit of NRQCD even at moderate energies.

Perturbative QCD factorization can also go beyond NRQCD fixed order calculations by resumming logarithms like $\ln (E_H^2/(2m_Q)^2)$ through evolution equations  \cite{Kang:2014tta}. We will solve the evolution equations to two loops to test the significance of evolution for these functions.

We begin in section \ref{sec:fact-summary} with a discussion of the factorization formalism of ref.\ \cite{Kang:2014tta}, identifying additional evanescent fragmentation functions associated with dimensional regularization.   We calculate the relevant short-distance coefficients for heavy quark pair production in association with gluons at NLO in section\ \ref{sec:lo-nlo}.   In section\ \ref{sec:comparison}, using the fragmentation functions of refs.\ \cite{Ma:2013yla,Ma:2014eja}, we compare our results to the asymptotic behavior of NRQCD calculations in the channels for which the $\mathcal{O}(\alpha^2\alpha_s^2)$ results are available and find agreement. Lastly, in section \ref{sec:numerical}, we study the approach to asymptotic behavior and evolution for the ${}^3S_1^{[1]}$ channel. We also show NLO pQCD predictions for different NRQCD channels at various center-of-mass (CM) energies, and include comparison to data from the Belle collaboration \cite{Pakhlov:2009nj}. 
 
\section{NLP factorization and fragmentation functions}
\label{sec:fact-summary}

As shown for LP in ref.\ \cite{Nayak:2005rt}  and NLP in ref.\ \cite{Kang:2014tta}, the perturbative short-distance coefficients of the factorized single-particle inclusive cross section, eq.\ (\ref{eq:pqcd_fac}) are sensitive only to the large lightlike momentum components of the parton(s) whose fragmentation produces the observed hadron. For NLP involving a heavy quark pair, the corresponding contributions to the cross section are convolutions in the longitudinal momenta of the heavy quark and antiquark of the pair between the short-distance coefficients and the fragmentation functions.  The short-distance coefficients and fragmentation functions describe the production of the pair, and its subsequent evolution into a heavy quarkonium, respectively. This NLP contribution is only one of many, but it is reasonable to assume that it dominates the class of NLP corrections simply because of the presence of the pair in heavy quarkonium production.

\subsection{Pinch surfaces and factorization}
In this subsection, we review the underlying structure that allows us to separate and normalize short-distance coefficients and fragmentation functions for NLP heavy quark pair production.  This requires the construction of the heavy quark pair fragmentation functions at the partonic level, which begins with diagrammatic analysis.

The observation that underlies factorization procedures is that long-distance contributions to cross sections for hard-scattering processes arise from well-defined regions in phase space and loop momenta. These regions are associated with so-called ``pinch surfaces" \cite{Sterman:1978bi,Libby:1978qf,Libby:1978bx} in the massless limit.   Each such region is associated with a specific power behavior in the high energy limit, LP, NLP and so forth.  The LP and NLP contributions in particular can be factorized into perturbatively-computable hard parts (short-distance coefficients), which determine the power behavior, and universal parton distributions for incoming hadrons and fragmentation functions for observed hadrons, which absorb mass and other long-distance dependence.   At large momentum transfer and to NLP, even the dependence on a heavy quark mass factorizes from the short-distance scattering \cite{Collins:1998rz}.   In this approximation,  LP and NLP contributions can be significantly simpler to determine than by derivation from full fixed-order calculations.

We recall first the procedure at LP.   A generic pinch surface contributing to LP is illustrated in figure\ \ref{fig:LPgluon}a for gluon fragmentation.  The momentum of the gluon passing the horizontal dashed line in the figure represents  a propagator near its mass shell.   In the region in question, the subdiagram below the cut describes the short-distance production of a nearly on-shell gluon from a hard scattering.  Subsequent fragmentation of the nearly on-shell gluon into a heavy quarkonium over long times is represented by the subdiagram above the line.    Every pinch surface that contributes to the cross section at LP takes this form (for a gluon or quark), with a short-distance piece, which is perturbatively calculable, and a non-perturbative but universal function. This is what makes factorization possible.   Given this factorization, it is possible to improve systematically the calculation of the short-distance coefficients, and as a result to determine systematically universal non-perturbative fragmentation functions by comparison to experiment or other non-perturbative input.   This program relies on the assumption that the calculation of the short-distance coefficients is by construction independent of long-time dynamics, and can be carried out in an infrared-regulated version of QCD.   In particular, for LP fragmentation, we normally carry out the computation of the short-distance coefficient in dimensionally-regularized QCD, taking the observed final-state hadron to be a parton itself.

\bef
 \center{\includegraphics[width=14cm]
 {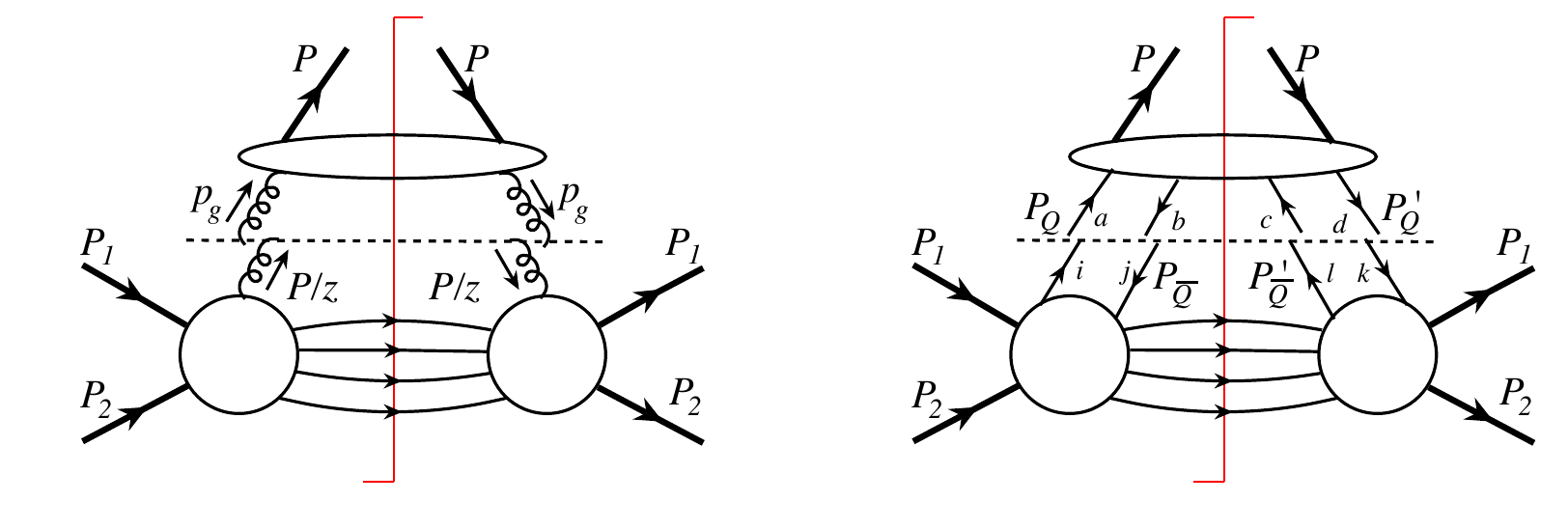}}
 \hbox{\hspace{1cm} (a) \hspace{6cm} (b)}
 \caption{\label{fig:LPgluon} 
 Pinch surfaces of the (a) single-parton leading power fragmentation and (b) heavy quark pair next-to-leading power fragmentation.}
\eef

Schematically, all contributions to all pinch surfaces at LP for the production of a hadron at high energy can be represented as 
\bea
d\sigma^{[\text{LP}]}_{e^+e^-\to H(P)} 
&\approx  \sum_a
\int_{x_H}^1 dz\,
\bigg[
{\cal H}_{e^+e^-\to {aX}}\left ( \frac{p}{z} \right ) {\cal T}_{a \to H(P)}(z,P)
\bigg]
d\Pi (P)\, ,
\label{eq:QQxsection_lp0}
\eea
where we recall from eq.\ (\ref{eq:momenta-defs}) that $p^\mu$ is the light-like projection of hadron momentum $P^\mu$.
The function ${\cal H}_{e^+e^-\to {aX}}$ represents all diagrammatic contributions to the production of parton $a$, below the dashed line in figure\ \ref{fig:LPgluon}a, and ${\cal T}_{a \to H(P)}$ those above.  Here for simplicity of notation, we take $0<x_H<1$ as the fractional momentum of the observed hadron H in $\rm e^+e^-$ annihilation, integrated over phase space $d\Pi(P)$. 

The starting form for our discussion for the contributions from generic NLP pinch surfaces  involving the production of a heavy quark pair at short distances \cite{Kang:2014tta}, illustrated by figure\ \ref{fig:LPgluon}b, is
\bea
d\sigma^{[\text{NLP}]}_{e^+e^-\to H(P)} 
&\approx 
\int_{x_H}^1 dz\, \int_0^1 du\, dv\ {\rm Tr}
\bigg[
{\cal H}_{e^+e^-\to [Q\bar{Q}]X}(p_Q, p_{\bar{Q}}, p'_Q, p'_{\bar{Q}})
\nonumber\\
&\  \hskip 1in
\times 
{\cal T}_{[Q\bar{Q}] \to H(P)}(z,u,v; P)
\bigg]
d\Pi (P)\nnu
& \approx  \ 
\int_{x_H}^1 dz\, \int_0^1 du\, dv\ 
\bigg[
{\cal H}_{ab,cd;ij,kl}(p_Q, p_{\bar{Q}}, p'_Q, p'_{\bar{Q}})
\nonumber\\
&\  \hskip 1in
\times 
{\cal T}^{ab,cd;ij,kl}(z,u,v; P)
\bigg]
d\Pi (P)
\,
\, .
\label{eq:QQxsection_nlp}
\eea
In the second expression we exhibit sums over two pairs of color indices ($ab,cd$) and two pairs of spin indices ($ij,kl$) linking the long distance part ${\cal T}$ and the short-distance part ${\cal H}$ in both the amplitude ($ab,ij$) and complex conjugate ($cd,kl$).    The form of eq.\ (\ref{eq:QQxsection_nlp}) is already suggestive of the factorized NLP cross section in eq.\ (\ref{eq:pqcd_fac}).   Bridging the gap between the two requires the identification of the dominant momentum dependence and the separation of color and spin traces between the long- and short-distance coefficients. These steps are described in ref.\ \cite{Kang:2014tta}, and we will recall them here for their detailed implementation at NLO. 

 Consider first the treatment of momentum flow.  As in eq.\ (\ref{eq:QQxsection_lp0}), the limit on the $z$ integral in eq.\ (\ref{eq:QQxsection_nlp}) refers to the production of a hadron integrated over phase space at fixed fractional momentum $x_H$, while the $u$ and $v$ integrals go from zero to one, representing the fraction of the heavy pair's momentum carried by the heavy quark in the amplitude and complex conjugate, respectively.

The hard functions ${\cal H} $ depend only on the light-like projections of the heavy quark and antiquark momenta on either side of the (vertical) final state cut.   Specifically, in terms of the parameters $u\dots \bar v$ introduced above, these momenta are given in terms of $p^\mu$ from eq.\ (\ref{eq:momenta-defs}) the light-cone projection of the final state momentum, $P^\mu$, by
\bea
p^\mu_Q =& 
\frac{u}{z}\, p^\mu\, ,  \quad p^\mu_{\bar{Q}} 
=
  \frac{\bar{u}}{z}\, p^\mu\, ,
\nonumber\\
p'^\mu_Q =& 
\frac{v}{z}\, p^\mu\, , 
\quad p'^\mu_{\bar{Q}} 
=
\frac{\bar{v}}{z}\, p^\mu\, .
\label{eq:hq-hat}
\eea
That is, in addition to the pair spin-color state $\kappa$, we must specify the momentum fractions $u$ and $v$ carried by the quarks in the amplitude and complex conjugate in the partonic final state. The factorized cross section will be a triple convolutions in $z$, $u$, and $v$.   

We now turn to the separation of the sums over color and spin indices.   Here, we follow the method of ref.\ \cite{Kang:2014tta}, modified for spin projections to accommodate dimensional regularization.
Quite generally, at fixed values of $u$ and $v$, we can expand the perturbative long-distance and short-distance functions into color singlet and octet matrices (components $ab$) times Dirac matrices (components $ij$). For example, for the short-distance function, $\cal H$, we find \cite{Kang:2014tta}
\bea
\left( {\cal H}_{e^+e^-\to [Q\bar{Q}]X}(p_Q, p_{\bar{Q}}, p'_Q, p'_{\bar{Q}}) \right)_{ab,ij;cd,kl} =& 
\delta_{ab}\, \delta_{cd} \sum_{I} \left( \Gamma^I \right)_{ij} \left( \Gamma^I \right)_{kl }\, {\cal H}_{1,I}\left(p_Q, p_{\bar{Q}}, p'_Q, p'_{\bar{Q}}\right) 
\nonumber\\
&\  \hspace{-25mm} + \ \sum_{A=1}^8\ \left( t_A\right)_{ab}\, \left( t_A\right)_{cd}\, \sum_{I} \left( \Gamma^I \right)_ {ij}\, \left( \Gamma^I \right)_ {kl} 
{\cal H}_{8,I}\left(p_Q, p_{\bar{Q}}, p'_Q, p'_{\bar{Q}} \right)
\, ,
\nonumber\\
\label{eq:H-expand}
\eea
where $\mathcal{H}_{a,I}$ are coefficients of the generators of spin state $I$ and color state $a$. The two Dirac matrices correspond to the amplitude and complex conjugate side of the short-distance coefficients and they are diagonal \cite{Kang:2014tta} for unpolarized initial and final states. The $t_A$ are SU(3) generators in fundamental representation. In four dimensions, the $\Gamma_I$ can be taken from the usual basis of Dirac matrices in four dimensions: 
\begin{equation}
\Gamma_I\ \in \ \left \{ {\bf 1},\gamma_\mu,\sigma_{\mu\nu},\gamma_\mu\gamma_5,\gamma_5 \right \}\, .
\label{eq:Fierz-set}
\end{equation}
 We will use the structure of eq.\ (\ref{eq:H-expand}) as a guide in the construction of factorizing projections in color and spin.

\subsection{Color projections}
The long- and short-distance parts in eq.\ (\ref{eq:QQxsection_nlp}) are connected by two heavy quark pairs, each of which can be in a singlet or octet state.   The projection in eq.\ (\ref{eq:QQxsection_nlp}) is standard, and the process may be represented explicitly by the relations
\bea
\sum_{a,b;c,d}\ =&\ \sum_{a,b;c,d} \sum_{a',b';c',d'}\, \delta_{aa'}\delta_{bb'}\delta_{cc'}\delta_{dd'}
\nonumber\\[2mm]
=&\ \sum_{a,b;c,d} \sum_{a',b';c',d'} \left [ \cc{C}_{ab,cd}^{[1]} \tilde{\cc{C}}_{a'b',c'd'}^{[1]} + \cc{C}_{ab,cd}^{[8]} \tilde{\cc{C}}_{a'b',c'd'}^{[8]} \right ]\, .
\label{eq:colorfierz}
\eea
Here we take the color projections $\cc{C}$  for the heavy quark pairs that enter ${\cal T}$ and isolate ones that give singlet or octet pair production.  Gauge invariance ensures that these projections are diagonal between the amplitude and complex conjugate.  Acting on $\cal T$ (above the dashed line), these are  \cite{Kang:2014tta}
\bea
\cc{C}_{ab,cd}^{[1]} =& \left[\f{\delta_{ab}}{\sqrt{N_c}}\right]\left[\f{\delta_{cd}}{\sqrt{N_c}}\right]\,,
\nnu
\cc{C}_{ab,cd}^{[8]} =& \f{1}{N_c^2-1}\sum_B [\sqrt{2}(t_B)_{ab}][\sqrt{2}(t_B)_{cd}]\, .
\label{eq:C-frag}
\eea
Correspondingly, projections $\tilde{\cc{C}}$ acting on the hard subdiagrams are
\bea
\tilde{\cc{C}}_{ab,cd}^{[1]} =& \left[\f{\delta_{ab}}{\sqrt{N_c}}\right]\left[\f{\delta_{cd}}{\sqrt{N_c}}\right]\, ,
\nnu
\tilde{\cc{C}}_{ab,cd}^{[8]} =& \sum_B [\sqrt{2}(t_B)_{ab}][\sqrt{2}(t_B)_{cd}]\ .
\label{eq:C-hard}
\eea
We now turn to the factorization of the spin degrees of freedom.

\subsection{Spin projections and dimensional regularization}
The spin analog of the color decomposition given in eq.\ (\ref{eq:colorfierz}) in four dimensions is a Fierz decomposition for each pair, with a normalization suitable to the basis in eq.\ (\ref{eq:Fierz-set}),
\bea
\sum_{i,j;k,l} = \sum_{i,j;k,l} \sum_{i',j';k',l'}\delta_{ii'} \delta_{jj'}\, \delta_{kk'}\, \delta_{ll'}  \ =\ \sum_{i,j;k,l} \sum_{i',j';k',l'} \sum_{I,J} \  \left ( \Gamma_I \right )_{ij} \;  \left (\Gamma^I \right )_{i'j'} \  \left ( \Gamma_J \right )_{kl} \left (\Gamma^J \right )_{k'l'}\, .
\label{eq:Fierz}
\eea
 We shall see below that in calculations of $\cal T$ and $\cal H$ using dimensional regularization, we need to expand the basis of Dirac structures to reflect the arbitrary number of dimensions involved. The regularized cross section will include terms that are nonvanishing resulting from the product of poles associated with collinear radiation, a long-time process, times short-distance factors that vanish when the number of dimensions is taken to four. The introduction of new, evanescent projections \cite{Buras:1998raa,Herrlich:1994kh} will enable us to identify such terms, and organize them appropriately.   In particular, this analysis will be necessary to reconcile NLP factorization at next-to-leading order with existing NRQCD calculations in selected channels, and to provide unambiguous definitions for short-distance coefficients for channels in which explicit calculations have not been carried out.   
  
 The full extended Dirac algebra in $D$ dimensions is given by linear combinations of the elements of the sets
\bea
\label{eq:gammafs}
 \left \{  \Gamma_I \right \} \ =&\ \left \{ {\bf 1},\gamma_\mu,\sigma_{\mu\nu},\gamma_\mu\gamma_5,\gamma_5, \, \left \{ \Gamma_{\hat{I}_j} \right \} \right \}
\nonumber\\[2mm]
\left \{ \Gamma_{\hat{I}_j} \right \}  \ =&\ \left \{ \Gamma_I\, \times \gamma^{\hat{m}_1} \times \dots \times \gamma^{\hat{m}_j} \right \}\, ,
\eea
where all of the (distinct) hatted indices $\hat{m}_1 \ne \hat{m}_2 \dots \ne \hat{m}_j$ are outside the usual four dimensions.  We assume that all our continued dimensions are spacelike and that the trace of any product of Dirac matrices with all different indices vanishes, including a single matrix. Even though we usually think of dimensional continuation as infinitesimal, as long as $\epsilon=2-D/2$ remains variable, there is no limit to the size of 
the algebra generated this way.  Nevertheless, the number of elements of the algebra that can be realized in a given calculation is finite, since there are no more Dirac matrices than the number of vertices plus the number of propagators along the fermion lines in any given diagram.

We will use the full set of elements of the sets above, $\{\Gamma_I\}$ including  $\{\Gamma_{\hat{I}_j}\}$ to realize an orthonormal basis with which to expand both $\cal T$ and $\cal H$.  As a result, the traces that link the long- and short-distance parts will be nonzero only when the Dirac structure of the two factors matches exactly for the two functions.   As we expand the short-distance part to higher orders, the expansion has more and more terms from $\{\Gamma_{\hat{I}_j}\}$.
For the project at hand, however, we only encounter three Dirac matrices with the evanescent projections.  This makes the calculations described below more manageable than they might otherwise have been.

To facilitate the discussion of the new evanescent projections for our NLO calculations, we introduce the notation
for any value of $\mu$ and $\nu$,
\bea
\gamma^{\mu\nu} =\frac{1}{2}\left( \gamma^\mu \gamma^\nu - \gamma^\nu \gamma^\mu\right)\, .
\label{eq:gamma-munu-def}
\eea
Of special interest for us in this calculation are the matrices with one and two hatted indices,
\bea
\gamma^{m\hat{n}}   \ =&\ \gamma^m \gamma^{\hat{n}}
\nonumber\\
\gamma^{\hat{m}\hat{n}}   \ =&\ \gamma^{\hat{m}} \gamma^{\hat{n}}  \,,
\eea
where $\hat{m} \neq \hat{n}$ again. 
From this point on, we will call states associated with these evanescent projections, eva$A$ and eva$B$, respectively.

As observed in ref.\ \cite{Kang:2014tta}, in the frame where the heavy quarkonium moves in the $z$-direction, as in eq.\ (\ref{eq:momenta-defs}), the leading contributions of the long-distance functions $\cc{T}$ are proportional to matrices in with large $+$ components dominate, i.e. terms projected out with $\Gamma_I \sim \sla{n}$. Therefore, in the limit $m_Q/E_H \to 0$, we require in eq.\ (\ref{eq:Fierz}) $\Gamma_I \sim \sla{n}$ above the cut and $\Gamma^I \sim \sla{p}$ below the cut. This is the case in arbitrary numbers of dimensions. The observed momentum $p^\mu$ is kept in four dimensions, so that the full set of NLP matrices below the cut is given by suitably normalized elements of the set
\bea
 \left\{ \rlap{p}{/}\,\, ,\, \rlap{p}{/}\, \gamma_5\, ,\, \rlap{p}{/} \gamma^{m\hat{n}}\, ,\,\rlap{p}{/}  \gamma^{\hat{m}\hat{n}}  \right \}\, ,
\label{eq:full-list}
\eea
in both for the amplitude and complex conjugate side. In the absence of polarizations in the initial state, the expansion of the hard scattering in terms of Dirac matrices is diagonal in these matrices, as anticipated in eq.\  (\ref{eq:H-expand}) above. This diagonality is then carried over to the general spin decomposition given by eq.\ (\ref{eq:Fierz}), which ensures that the fragmentation functions are also diagonal in the projections. New evanescent fragmentation functions based on these Dirac matrices must also be added to achieve exact NLP factorization in dimensional regularization. The first two matrices in eq.\ (\ref{eq:full-list}), which appear with both singlet and octet color, correspond to longitudinal and scalar polarization configurations for the pair. They are referred to as vector $(v)$ and axial $(a)$ projections. We can think of the additional matrices with hatted indices as extensions of  $\rlap{p}{/}\gamma_5$ into $D$ dimensions.   For us, $\gamma_5$ remains in four dimensions.   Of course, for $D=4$, only the first two of these structures are necessary, and they appear in the Fierz identity, eq.\ (\ref{eq:Fierz}).
Transverse projections are also possible, of course, but are associated with even powers of Dirac matrices, and vanish at NLP for the gluon-associated processes we consider \cite{Kang:2014tta}.

The resulting, normalized spin projection operators for the $v$, $a$, eva$A$, and eva$B$ fragmentation functions are given by 
\bea
\mathcal{P}^{(v)}(p)_{ij,lk} =& \frac{1}{4p\cdot n}\left(\gamma \cdot n\right)_{ij} \, \frac{1}{4p\cdot n}\left(\gamma \cdot n\right)_{lk}\,,\nnu
\mathcal{P}^{(a)}(p)_{ij,lk} =&  \frac{1}{4p\cdot n}\left(\gamma \cdot n \gamma_5\right)_{ij} \, \frac{1}{4p\cdot n}\left(\gamma \cdot n \gamma_5\right)_{lk}\nnu
=&\frac{i}{\sqrt{2}} \frac{1}{4p\cdot n}\left(\gamma \cdot n \gamma^{mn}\right)_{ij} \,\frac{i}{\sqrt{2}}  \frac{1}{4p\cdot n}\left(\gamma \cdot n \gamma^{mn}\right)_{lk}\,,\nnu
\mathcal{P}^{(\mathrm{eva}A)}(p)_{ij,lk} =&\frac{2}{D-4}\frac{i}{\sqrt{2}} \frac{1}{4p\cdot n}\left(\gamma \cdot n \gamma^{m\hat{n}}\right)_{ij} \,\frac{i}{\sqrt{2}}  \frac{1}{4p\cdot n}\left(\gamma \cdot n \gamma^{m\hat{n}}\right)_{lk}\,,\nnu
\mathcal{P}^{(\mathrm{eva}B)}(p)_{ij,lk} =&\frac{2}{(D-4)(D-5)}\frac{i}{\sqrt{2}} \frac{1}{4p\cdot n}\left(\gamma \cdot n \gamma^{\hat{m}\hat{n}}\right)_{ij} \,\frac{i}{\sqrt{2}}  \frac{1}{4p\cdot n}\left(\gamma \cdot n \gamma^{\hat{m}\hat{n}}\right)_{lk},
\label{eq:T-spin-projections}
\eea
where repeated indices are summed. 

The corresponding projection operators for hard, short-distance coefficients are given by
\bea
\tilde{\mathcal{P}}^{(v)}(p)_{ji,kl} =& \left(\gamma \cdot p\right)_{ji} \,\left(\gamma \cdot p\right)_{kl}\,,\nnu
\tilde{\mathcal{P}}^{(a)}(p)_{ji,kl} =& \left(\gamma \cdot p \gamma_5\right)_{ji} \,\left(\gamma \cdot p \gamma_5\right)_{kl}\nnu
=&\frac{i}{\sqrt{2}} \left(\gamma \cdot p \gamma^{mn}\right)_{ji} \,\frac{i}{\sqrt{2}} \left(\gamma \cdot p \gamma^{mn}\right)_{kl}\,,\nnu
\tilde{\mathcal{P}}^{(\mathrm{eva}A)}(p)_{ji,kl} =&\frac{i}{\sqrt{2}} \left(\gamma \cdot p \gamma^{m\hat{n}}\right)_{ji} \,\frac{i}{\sqrt{2}} \left(\gamma \cdot p \gamma^{m\hat{n}}\right)_{kl}\,,\nnu
\tilde{\mathcal{P}}^{(\mathrm{eva}B)}(p)_{ji,kl} =&\frac{i}{\sqrt{2}} \left(\gamma \cdot p \gamma^{\hat{m}\hat{n}}\right)_{ji} \,\frac{i}{\sqrt{2}} \left(\gamma \cdot p \gamma^{\hat{m}\hat{n}}\right)_{kl}\,. 
\label{eq:P-tilde-H}
\eea

In principle, there are separate evanescent functions for each combination of hatted indices, but for four-dimensional final spin and momentum states, they are all related by rotations. Using rotational invariance in dimensions beyond four, the functions are the same for every value of the hatted indices.   As a result we sum over all hatted indices, dividing by the number of extra dimensions.   Of course, this counting is only possible in integer numbers of dimensions, but it has an analytic continuation to any complex value of $\epsilon=2-D/2$.

The overall factors for fragmentation projections in eq.\ (\ref{eq:T-spin-projections}) eva$A$ and eva$B$ come with overall counting normalizations (for the collinear fragmentation functions)
\bea
\frac{2}{D-4} \qquad \text{and}\qquad \frac{2}{(D-4)(D-5)}\,,
\eea
respectively. This counting normalization ensures the orthonormality condition
\bea
\sum_{ijlk} \mathcal{P}_{ij,lk}^{(s)}(p)\tilde{\mathcal{P}}_{ji,kl}^{(s')}(p) = \delta^{ss'}
\eea
with $s,s' = v,a,\mathrm{eva}A,\mathrm{eva}B$. As long as such orthonormality condition is satisfied, the exact normalizations of $\mathcal{P}$ and $\tilde{\mathcal{P}}$ are not important. However, since both $\frac{2}{D-4}$ and $\frac{2}{(D-4)(D-5)}$ are $\sim \frac{1}{\epsilon}$, depending on whether we choose to put this overall normalization with the hard projections or fragmentation projections we change the order of $\epsilon$ in the hard coefficients or fragmentation functions. We choose to include these counting factors in the projectors for the fragmentation functions, and use $\overline{\text{MS}}$ scheme for the evanescent as well as four-dimensional fragmentation functions. This compensates for the suppression induced by summing over the number of dimensions beyond four in eq.\ (\ref{eq:T-spin-projections}). For fixed hatted indices and $\epsilon$ independent normalizations, the corresponding fragmentation functions are proportional to $1/\epsilon$, just as for four dimensional projections. Of course, our final results are scheme independent, but our choice is physically consistent with the above remark that evanescent contributions come from order $\epsilon$ short-distance parts convolved with poles in long-distance fragmentation functions.

With these normalizations, we can generalize the Fierz identity in four dimensions to
\bea
\label{eq:spinfierz}
\sum_{i,j;k,l} = \sum_{i,j;k,l} \sum_{i',j';k',l'}\delta_{ii'} \delta_{jj'}\, \delta_{ll'} \delta_{kk'} = \sum_{i,j;k,l} \sum_{i',j';k',l'}\sum_{I=v,a,\mathrm{eva}A,\mathrm{eva}B} \mathcal{P}^{(I)}(p)_{ij,lk} \tilde{\mathcal{P}}^{(I)}(p)_{j'i',k'l'} \ +\ R\, ,
\eea
where the primed indices are contracted with the hard scattering, and the unprimed with the fragmentation functions, and where the remainder, $R$, contributes only beyond NLP. $R$ includes the transverse projections, which may also be NLP for other processes.   Again, we can identify a finite list of projections because we are working at finite order in the hard scattering at NLP.  Now we can separate both color and spin traces in the integrand of eq.\ (\ref{eq:QQxsection_nlp}), using eq.\ (\ref{eq:colorfierz}) and eq.\ (\ref{eq:spinfierz}), respectively, which, with standard approximations on loop momenta, give us the NLP part of the factorization found in eq.\ (\ref{eq:pqcd_fac}) \cite{Kang:2014tta}.

\subsection{Operator definitions for partonic fragmentation functions}
Each pair of projections, eqs.\ (\ref{eq:C-frag}) for color and (\ref{eq:T-spin-projections}) for spin, defines a fragmentation function.   The result is given above in eq.\ (\ref{eq:calD-def}).
Compared to ref.\ \cite{Kang:2014tta}, we now include in the set of projections the full list of $D$-dimensional Dirac structures in eq.\ (\ref{eq:T-spin-projections}).

As in the single-parton case, we will compute the short-distance coefficients for the heavy quark pair states from IR regulated, $D$-dimensional partonic cross sections, using perturbative and regulated partonic fragmentation functions.  In these calculations, the heavy quark mass is set to zero. In contrast to the single-parton case, however, although the total momentum of the pair must be the same in the amplitude and complex conjugate of figure\ \ref{fig:LPgluon}b,  the relative momenta  of the quark and antiquark within the two pairs need not be the same.    This simply means that there can be different histories on how a heavy quark pair reaches the final heavy quarkonium state. The partonic implementation of the pair-to-hadron fragmentation function, eq.\ (\ref{eq:calD-def}), is therefore slightly different than  for single-parton fragmentation.  

 To compute partonic short-distance coefficients, we define fragmentation functions that can be directly applied to products of amplitudes in the infrared-regulated theory. These partonic fragmentation functions depend on fractional momenta of the heavy quarks in the final state.   They are given by, again suppressing the gauge links, 
\bea
&\  {\cal D}_{[Q\bar{Q}(\kappa)]\to [Q \bar Q](\kappa')}(z,u,v,u',v')
=
\nonumber\\
&\  \hskip 0.45in
 \int \frac{p^+ dy^-}{2\pi}\, {\rm e}^{-i(p^+/z)y^-}
\int \frac{p^+ dy_1^-}{2\pi}\, {\rm e}^{i(p^+/z)(1-v)y_1^-}
\int \frac{p^+ dy_2^-}{2\pi}\, {\rm e}^{-i(p^+/z)(1-u)y_2^-}
\nonumber \\
&\  \hskip 1 in
\times \,
{\cal P}^{(s_\kappa)}_{ij,lk}(p)\, {\cal C}^{[I_\kappa]}_{ab,cd}\ \sum_X
\langle 0|\overline{\psi}_{c,l}(y_1^-)\psi_{d,k}(0)|| [Q \bar Q](\kappa',v')X\rangle
\nonumber\\
&\  \hspace{2.5 in} \times\,
\langle  [Q \bar Q](\kappa',u')X| \overline{\psi}_{a,i}(y^-)\psi_{b,j}(y^- + y_2^-)|0\rangle \, ,
\nonumber\\
\label{eq:calD-partonic-def}
\eea 
where $p^+$ is the lightlike projection of the heavy quark pair
momentum, and in both final states, $p_Q+p_{\bar{Q}}=p$. In these matrix elements, the ``heavy'' quarks are treated as massless, since they are designed to match the long-time behavior of the products of amplitudes in the infrared-regulated theory.
  
The produced pair(s) $Q \bar Q (\kappa',u')$ are labelled explicitly by the relevant heavy quark momentum fraction.   In each of these Fock states we take the final heavy quark and antiquark momenta to be lightlike and parallel to the projection of the heavy quark pair's momentum, $p$
\bea
&\  p_Q\ =\ u' p \, , \quad p_{\bar{Q}}\ =\ \bar u' p\ {\rm (amplitude)},
\nonumber\\
&\  p_Q\ =\ v' p\, , \quad p_{\bar{Q}}\ =\ \bar v' p\ {\rm (conjugate)}.
\eea
Normalizing the quark-pair states to the projection operators in eq.\ (\ref{eq:T-spin-projections}) we have at LO,
\bea
\label{eq:LOfrag}
{\cal D}_{[Q\bar{Q}(\kappa)]\to [Q\bar{Q}(\kappa')]}^{(0)}(z,u,v,u',v') =  \delta_{\kappa \kappa'}\,\delta(1-z)\,\delta(u-u')\,\delta(v-v')\,.
\eea
This simple result makes the relation between the LO short-distance coefficients and the Born cross section eq.~(\ref{eq:pqcd_fac}) direct, as we will see below.

 The partonic fragmentation functions in eq.\ (\ref{eq:calD-partonic-def}) were employed in ref.\ \cite{Kang:2014tta} to identify evolution kernels.   Here, we use them to compute partonic fragmentation functions at order $\as$, and it is useful to have an explicit definition.   In principle, it is possible to combine these off-diagonal fragmentation functions into a more direct analog of the hadronic functions in eq.\ (\ref{eq:calD-def}).   This would be carried out by combining linear combinations of states with fixed $u'$ and $v'$ into some ``pure" partonic final state characterized only by the total momentum of the pair.   This is not necessary for our calculations here, however, so we will not develop the formalism in this direction.

\subsection{Partonic fragmentation functions at order $\as$}
\bef
 \center{\includegraphics[width=16cm]
 {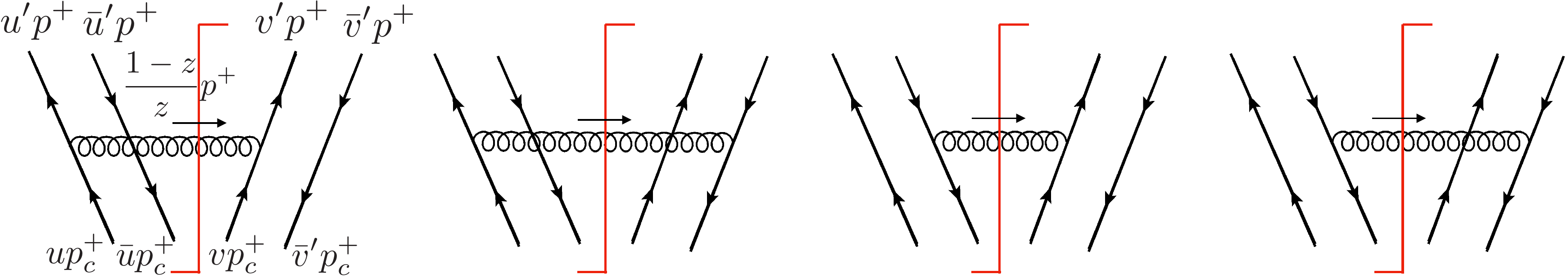}}
 \caption{\label{fig:realFF} Real emission diagrams for computing partonic fragmentation functions at order $\alpha_s$ for light-cone gauge. We label the flow of $+$ components of momentum to make $z,u,v,u',v'$ variable dependence explicit. For the full momentum dependence and details of the computation, see ref.\ \cite{Kang:2014tta}.}
\eef

At order $\alpha_s$, the partonic fragmentation functions for the four dimensional basis were calculated in ref.\ \cite{Kang:2014tta} for the four-dimensional projections $s=v,a$ of eq.\ (\ref{eq:T-spin-projections}).  
Here, we give the set of partonic fragmentation functions needed for the factorized NLP cross section at $\mathcal{O}(\as)$.  The only difference is
in the Dirac projections, and we do not give the details of the computation here. Recalling that our final states are always four-dimensional, we only need fragmentation functions
with $\kappa'=v,a$ in color octet or singlet for NLO calculations of our process.   The
$\rm \overline MS$ scheme results can be written as
\bea
\label{eq:partDMS}
\cc{D}^{(1),\overline{\text{MS}}}_{[Q\bar{Q}(\kappa)]\to Q\bar{Q}(\kappa')]}(z,u,v,u',v',\mu) = -\left(\f{1}{\varepsilon} + \ln 4\pi e^{-\gamma_E}\right)\f{\alpha_s}{2\pi}P_{\kappa \to \kappa'}
\eea
where the kernels $P_{\kappa\to\kappa'}$ for the non-evanescent states are given in the appendix of \cite{Kang:2014tta}.

We extend the results of \cite{Kang:2014tta} with the relevant evanescent fragmentation functions of our process, namely $\kappa = \text{eva}A[8],\,\text{eva}B[8]$ and $\kappa' = v,a$ in color singlet or octet state. Since these are all off-diagonal splitting functions, at $\mathcal{O}(\as)$, they are given by the real emission diagrams in figure \ref{fig:realFF} for light-cone gauge. We find
\bea
\label{eq:P-1-evaAv1}
P_{\text{eva}A[8]\to v[1]} =& \left[\frac{1}{2N_c}\right] \frac{z}{(1-z)} S_- \Delta_-^{[1]}\,, \\
P_{\text{eva}A[8]\to a[1]} =&  \left[\frac{1}{2N_c}\right] \frac{z}{(1-z)} S_- \Delta_+^{[1]}\,,\\
P_{\text{eva}A[8]\to v[8]} =& \left[\frac{1}{2N_c}\right] \frac{z}{(1-z)} S_- \Delta_-^{[8]} \,,\\
P_{\text{eva}A[8]\to a[8]} =&  \left[\frac{1}{2N_c}\right] \frac{z}{(1-z)} S_- \Delta_+^{[8]}\,,\\
P_{\text{eva}B[8]\to v[1]} =& \left[\frac{1}{2N_c}\right] \frac{z}{2(1-z)} S_- \Delta_-^{[1]}\,,\\
P_{\text{eva}B[8]\to a[1]} =& 0\,,\\
P_{\text{eva}B[8]\to v[8]} =& \left[\frac{1}{2N_c}\right] \frac{z}{2(1-z)} S_- \Delta_-^{[8]}\,,\\
P_{\text{eva}B[8]\to a[8]} =& 0\,,
\label{eq:P-1-evaBa8}
\eea
where the  functions $S_\pm$ and $\Delta_\pm^{[1,8]}$ are defined by
\bea
\label{eq:partFDeltas}
S_\pm =\, & \left (\f{u}{u'}\pm \f{\bar{u}}{\bar{u}'} \right )\, \left ( \f{v}{v'}\pm \f{\bar{v}}{\bar{v}'} \right )\\[2mm]
\label{eq:Dpm1}
\Delta_\pm^{[1]} =\, & [\delta(u-z u') \pm \delta(\bar{u}-z\bar{u}')][\delta(v-zv') \pm \delta(\bar{v}-z\bar{v}')]\\[2mm]
\label{eq:Dpm8}
\Delta_\pm^{[8]} =\, & \bigg\{(N_c^2-2)[\delta(u-z u')\delta(v-zv') + \delta(\bar{u}-z\bar{u}')\delta(\bar{v}-z\bar{v}')] \nnu[2mm]
&\hspace{1.1cm}\mp 2[\delta(u-z u')\delta(v-zv') + \delta(\bar{u}-z\bar{u}')\delta(\bar{v}-z\bar{v}')]\bigg\}\,.
\eea

As we will demonstrate below, these evanescent fragmentation functions are crucial in our factorization to maintain scheme independence and agree with NRQCD calculation.

\section{Cross sections and coefficient functions}
\label{sec:lo-nlo}
With the order $\as$ pair fragmentation functions in hand, we are ready to derive the short-distance coefficients at LO and NLO.  We set the stage by exhibiting the short-distance coefficients in terms of partonic cross sections.   We then compute the partonic cross sections for the states of interest up to NLO. Finally, we use the partonic fragmentation functions found in section\ \ref{sec:fact-summary} to derive  expressions for the NLO short-distance coefficients.

\subsection{Coefficient functions from cross sections}
To compute the NLP short-distance coefficients $d\hat{\sigma}_{e^+e^- \to [Q\bar{Q}(\kappa)]}^{(n)}$, we replace the heavy quarkonium $H$ of eq.~(\ref{eq:QCDfact}) by $[Q\bar{Q}(\kappa)]$ and compute the left-hand and right-hand sides to $n$-th order in $\alpha_s$.    
We outline the procedure, in schematic notation, for the process under consideration.

For an arbitrary differential cross section, we write in the condensed notation of eq.\ (\ref{eq:QCDfact}),
\bea
d\sigma_{e^+e^- \to [Q\bar{Q}(\kappa)]}(p,u,v) =& \sum_{f} d\hat{\sigma}_{e^+e^- \to f}(p/z)\otimes_z D_{f\to [Q\bar{Q}(\kappa)]}(z,u,v)\nnu
&\hspace{-2cm}+\sum_{[Q\bar{Q}(\kappa')]} d\hat{\sigma}_{e^+e^- \to [Q\bar{Q}(\kappa')]}(p/z,u',v')\otimes_{z;u',v'} \cc{D}_{[Q\bar{Q}(\kappa')]\to [Q\bar{Q}(\kappa)]}(z,u',v',u,v)\,,
\label{eq:epem-nlp-fact}
\eea
where $p$ is the final heavy quark pair's momentum. In this expression, the arguments of the functions and the corresponding convolution symbols are taken as in eq.\ (\ref{eq:pqcd_fac}). We will suppress the explicit arguments of fractional momenta below.
At lowest order for the gluon-associated production process shown in figure \ref{eeLO}, at $\cc{O}(\alpha_s)$ and with the final state $[Q\bar{Q}(\kappa)]g$, only the second term contributes,
\bea
d\sigma^{(1)}_{e^+e^- \to [Q\bar{Q}(\kappa)]g} \ =\ \sum_{[Q\bar{Q}(\kappa')]} d\hat{\sigma}^{(1)}_{e^+e^- \to [Q\bar{Q}(\kappa')]g}\otimes_{z;u',v'} \cc{D}_{[Q\bar{Q}(\kappa')]\to [Q\bar{Q}(\kappa)]}^{(0)}\, .
\eea
Then, using eq.\ (\ref{eq:LOfrag}) for the zeroth order fragmentation functions, we arrive at
\bea
d\hat{\sigma}^{(1)}_{e^+e^- \to [Q\bar{Q}(\kappa)]g} = d\sigma^{(1)}_{e^+e^- \to [Q\bar{Q}(\kappa)]g} \, .
\label{eq:hat-sgma-1}
\eea
The short-distance coefficient is therefore equal to the partonic cross section at lowest order.  Of course, for LO leptonic annihilation, $\kappa$ is always an octet configuration.

Evaluating eq.\ (\ref{eq:epem-nlp-fact}) at $\cc{O}(\alpha_s^2)$ in the same way, but with two gluons in the final state, gives
\bea
d\sigma^{(2)}_{e^+e^- \to [Q\bar{Q}(\kappa)]gg} \ =\ &\sum_{[Q\bar{Q}(\kappa')]} d\hat{\sigma}^{(1)}_{e^+e^- \to [Q\bar{Q}(\kappa')]g}\otimes_{z;u',v'} \cc{D}_{[Q\bar{Q}(\kappa')]\to [Q\bar{Q}(\kappa)]g}^{(1)} \nnu
+&\sum_{[Q\bar{Q}(\kappa')]} d\hat{\sigma}^{(2)}_{e^+e^- \to [Q\bar{Q}(\kappa')]gg}\otimes_{z;u',v'} \cc{D}_{[Q\bar{Q}(\kappa')]\to [Q\bar{Q}(\kappa)]}^{(0)}\, .
\eea
Note again that we only compute the real diagrams at $\cc{O}(\alpha_s^2)$ as we focus on the region $x_H$ away from $1$. Then using eq.\ (\ref{eq:LOfrag}) for ${\cal D}^{(0)}$, we solve directly for the short-distance coefficient at order $\as^2$,
\bea
\label{eq:order2}
d\hat{\sigma}^{(2)}_{e^+e^- \to [Q\bar{Q}(\kappa)]gg} = d\sigma^{(2)}_{e^+e^- \to [Q\bar{Q}(\kappa)]gg} - \sum_{[Q\bar{Q}(\kappa')]} d\hat{\sigma}^{(1)}_{e^+e^- \to [Q\bar{Q}(\kappa')]g}\otimes_{z;u',v'} \cc{D}_{[Q\bar{Q}(\kappa')]\to [Q\bar{Q}(\kappa)]g}^{(1)}\, ,
\eea
where now $\kappa$ can be singlet or octet.\footnote{From this point on, we omit gluons in the final states for convenience.
Also, when there is no ambiguity, we will often simply use $\kappa$ in place of $Q\bar{Q}(\kappa)$.} The subtraction term on the right-hand side removes the long-distance behavior in the $\mathcal{O}(\as^2)$ cross section, which is due to the collinear emission of a gluon by the pair, a process that requires times
that are  large compared to $1/E_H$.

\subsection{Partonic projections and phase space}
We next establish the notation and projections necessary to calculate the partonic cross sections on the right-hand side of eqs.\ (\ref{eq:hat-sgma-1}) and (\ref{eq:order2}).  As in the discussion of partonic fragmentation functions, eq.\ (\ref{eq:calD-partonic-def}), we will use the projection operators of eq.\ (\ref{eq:P-tilde-H}) to define our partonic cross sections, 
\bea
d\sigma_\kappa^{(m)} (p,u,v) \equiv & d\sigma^{(m)}_{e^+e^-\to [Q\bar Q](\kappa)X}(p,u,v)\ \nnu
=&\  \frac{1}{2Q^2} \bar M^{*,(m)}_{e^+e^-\to [Q\bar Q](\kappa)X}(p,v)\, \bar M^{(m)}_{e^+e^-\to [Q\bar Q](\kappa)X}(p,u)\, \,d\Pi_{m+1}\,,
\nonumber\\
 \equiv &  \frac{1}{2Q^2} \, | {\cal M}_\kappa^{(m)}|^2(p,u,v)\,d\Pi_{m+1}\,,
 \label{eq:M-notation}
\eea
where the superscript $m = 1$ or $2$ denotes the order of $\alpha_s$, which is also equal to the number of gluons radiated, corresponding to the $m+1$-particle phase space factor $d\Pi_{m+1}$. Here, the amplitudes $\bar M$ are computed perturbatively, by stripping the final-state spinors for the pair and replacing them by a combination of the spin and color projections shown above.  
 In explicit calculations to produce spin state $s$, and color state $I$, we thus make the replacement
\bea
\bar u\left( uP,\lambda_Q\right)_{ib} v\left(\bar uP,\lambda_{\bar Q}\right)_{ja} \ \left[ \bar u\left( vP,\lambda_Q\right)_{kc} v\left(\bar vP,\lambda_{\bar Q}\right)_{ld} \right]^* \ 
\Rightarrow\   \tilde{\cc{P}}^{(s)}(p)_{ji,kl}\,  \tilde {\cal C}^{I}_{ab,cd} \, , 
\eea
for $s=v,a,\text{eva}A,\text{eva}B$, and similarly for singlet and octet color states.  The third line of eq.\ (\ref{eq:M-notation}) defines the ``squared" amplitude in this context.  
As for the perturbative computation of partonic fragmentation functions,
only the total momentum of the pair is the same in the amplitude and complex conjugate.

Finally, Lorentz-invariant phase space in eq.\ (\ref{eq:M-notation}) is given in $D=4-2\epsilon$ dimensions, not including the symmetry factors, by
\bea
\label{eq:phasespace}
d\Pi_n \equiv \prod_{\text{final states,i}}^n \frac{d^{D-1}p_i}{(2\pi)^{D-1}2E_i} (2\pi)^D \delta^D(k_1 + k_2 - \Sigma p_i)\, ,
\eea
where $k_i = (E_{k_i},\vec{k}_i)$ and $p_i = (E_i,\vec{p}_i)$ label the momenta of the incoming particles and the outgoing particles, respectively. For the LO diagrams of figure \ref{eeLO} and the NLO real diagrams in figure \ref{eeNLO}, eq.~(\ref{eq:M-notation}) 
then becomes
\bea
d\sigma^{(1)}_\kappa = \frac{1}{2Q^2} |\mathcal{M}^{(1)}_\kappa|^2 d\Pi_2\, ,
\label{eq:LOsigma}
\eea
and
\bea
\label{eq:NLOsigma}
d\sigma^{(2)}_\kappa = \frac{1}{2Q^2} |\mathcal{M}^{(2)}_\kappa|^2 d\Pi_3\left(\frac{1}{2}\right)\,,
\eea
respectively. The extra $1/2$ in eq.\ (\ref{eq:NLOsigma}) is the symmetry factor associated with the two gluons in the final state.
\bef
 \center{\includegraphics[width=8cm]
 {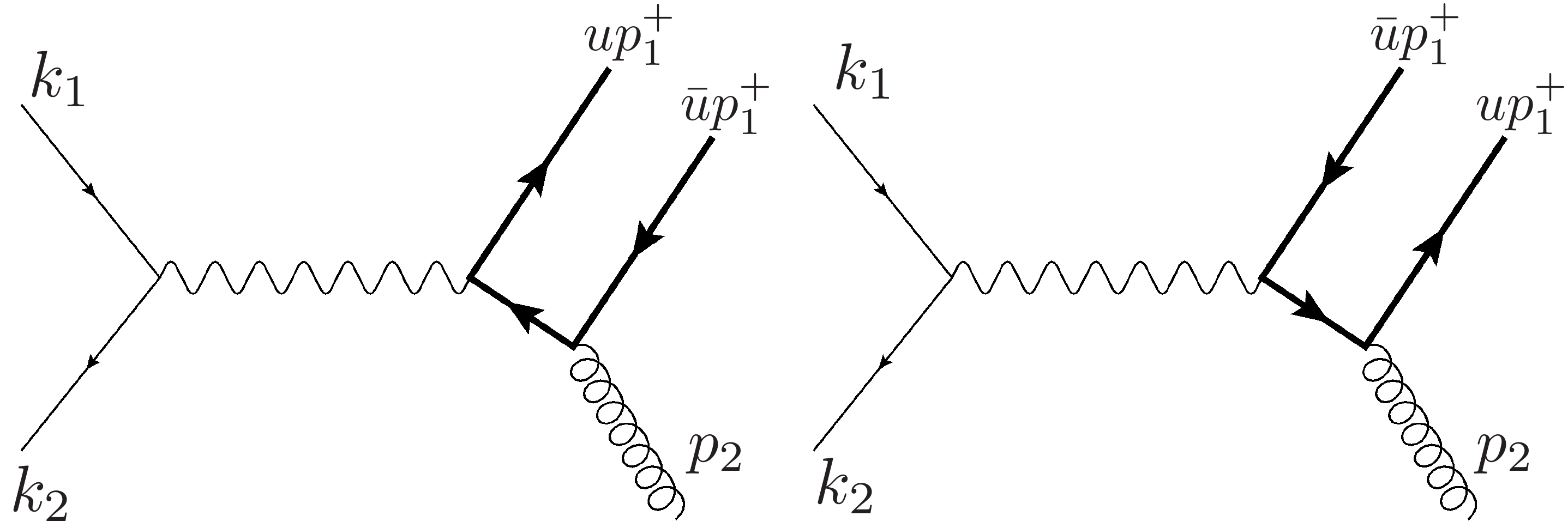}}
 \caption{\label{eeLO}Leading order Feynman diagrams for $e^+e^- \to [Q\bar{Q}(\kappa)]g$.}
\eef
\bef
 \center{\includegraphics[width=8cm]
 {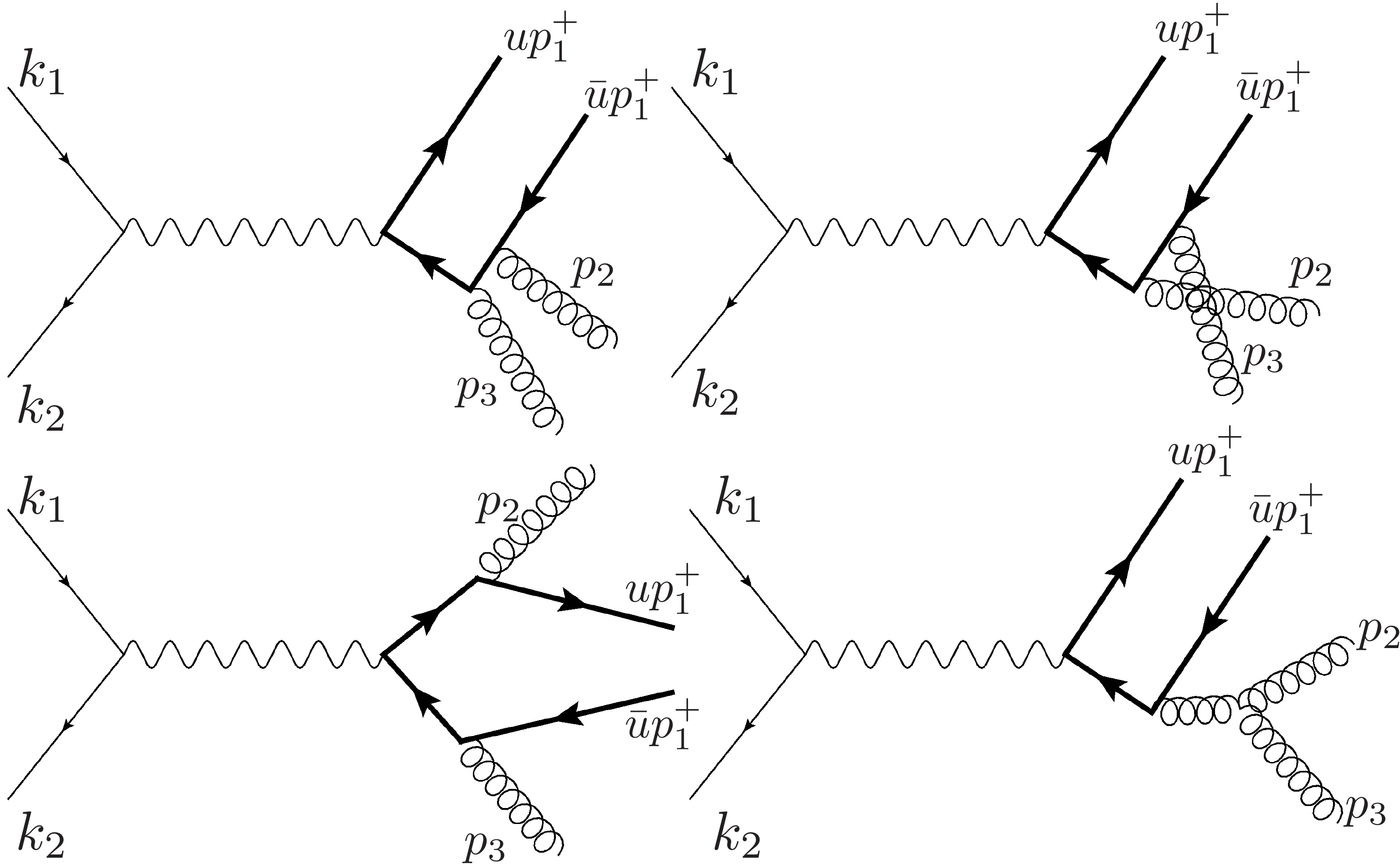}}
 \caption{\label{eeNLO}Four of the eight next-leading order Feynman diagrams for $e^+e^- \to [Q\bar{Q}(\kappa)]gg$ at $\cc{O}(\alpha_s^2)$. The other four are from flipping the arrows on the fermion lines. Virtual diagrams only contribute to $\delta(1-x_H)$ terms.}
\eef

We parameterize phase space in terms of fractional energies, labelling the pair's total energy as 
$E_1$. In particular, for three-particle phase space we use
\bea
x_i \ =\ \frac{2E_i}{Q}\, , \qquad \qquad
\sum_{i=1}^3 x_i\ =\ 2\, .
\eea
For an unpolarized initial state, the $D$-dimensional phase space factors can be integrated over angles to give 
\bea
\label{eq:Pi2}
d\Pi_2 =&\left(\frac{4\pi}{Q^2}\right)^\epsilon \frac{\Gamma(1-\epsilon)}{\Gamma(2-2\epsilon)} \frac{1}{8\pi} dx_1 \delta(1-x_1)\,,\\
d\Pi_3 =&\frac{Q^2}{16(2\pi)^3}\left(\frac{4\pi}{Q^2}\right)^{2\epsilon}\frac{1}{\Gamma(2-2\epsilon)}\,y^{-\epsilon}\,(1-y)^{-\epsilon}\,(1-x_1)^{-\epsilon}\,x_1^{1-2\epsilon}\,dx_1\,dy\,,
\label{eq:Pi3}
\eea
where we define
\bea
\label{eq:ydef}
 y \ =\ \frac{1-x_2}{x_1}\, , \quad 
 0 \ \leq \ y\ \leq 1\, .
\eea
The limits $y=0$ and $y=1$ correspond to one of the two final state gluons in figure \ref{eeNLO} carrying away maximal energy, i.e. $x_2 = 1$ and $x_3 = 1$, respectively. This gives the configuration in which the other gluon becomes collinear to the observed heavy quark pair, giving rise to collinear singularities. We now turn to the calculation of matrix elements and cross sections.

\subsection{Dirac traces and cross sections}

We continue by separating the  squared S-matrix into the leptonic and hadronic tensors as
\bea
|\mathcal{M}^{(m)}_\kappa|^2 = L_{\mu\nu} H^{\mu\nu,(m)}_{\kappa}\,,
\label{eq:M2-LH}
\eea
where the superscripts, $(m)$, indicate the perturbative order in the strong coupling.   Here, the leptonic part, taken at lowest order in QED coupling with spin averaging, given in terms of electron and positron momenta $k_i$ is
\bea
L_{\mu\nu} = \frac{e^2}{Q^4}({k_1}_\mu {k_2}_\nu + {k_1}_\nu {k_2}_\mu - g_{\mu\nu}\,k_1\cdot k_2)\,.
\label{eq:L-LO}
\eea
As noted above, we integrate over angles at fixed fractional momentum, $x_1$\footnote{Technically, the cross section is also differential in $u$ and $v$ as discussed after eq.\ (\ref{eq:M-notation}).}.
For our $x_1$ distribution, using current conservation, $q_\mu H^{\mu\nu}=0$, with $q^\mu = (k_1+k_2)^\mu$, we can simplify the hadronic tensor as
\bea
H^{\mu\nu,(m)}_{\kappa} =& (q^\mu q^\nu - q^2 g^{\mu\nu}) H_\kappa^{(m)}\,.
\eea

We find for quarks with fractional change $e_Q$,
\bea
H_\kappa^{(m)} = & \frac{-g_{\mu\nu}H^{\mu\nu,(m)}_{\kappa}}{(3-2\epsilon)q^2}
\nonumber\\
\equiv & \frac{e^2 e_Q^2}{(3-2\epsilon)q^2} \left(g^{2}\mu^{2\epsilon}\right)^mF_{\kappa}^{(m)}(x_1,y,u,v,\epsilon) \, ,
\label{eq:H-F-def}
\eea
where, in the second expression, we have factored out the scale, $\mu$, and coupling dependence from $-g_{\mu\nu}H^{\mu\nu,(m)}$ to define $F_\kappa^{(m)}$.

We now combine the leptonic and hadronic tensors from eq.\ (\ref{eq:L-LO}) and (\ref{eq:H-F-def}), respectively, in the squared matrix element in eq.\ (\ref{eq:M2-LH}). These results are combined with two- and three-particle phase space, eqs.\ (\ref{eq:Pi2}) and \ (\ref{eq:Pi3}), in eqs.\ (\ref{eq:LOsigma}) and (\ref{eq:NLOsigma}) to derive expressions for the two- and three-particle cross sections,

\bea\
\label{eq:hatsig1}
\frac{d\sigma^{(1)}_\kappa}{dx_1} =&\, \sigma_0 \frac{\alpha_s e_Q^2}{Q^2} \pi \, F_{\kappa}^{(1)}(x_1,y,u,v,\epsilon)\,\delta(1-x_1)\\
\label{eq:hatsig2}
\frac{d\sigma^{(2)}_\kappa}{dx_1} =&\, \sigma_0\ \frac{\alpha_s^2 e_Q^2}{\Gamma(1-\epsilon)}  \frac{1}{8} \left(\frac{4\pi\mu^2}{Q^2 x_1^2 (1-x_1)}\right)^\epsilon  \int_0^1 dy\,F_{\kappa}^{(2)}(x_1,y,u,v,\epsilon) \, y^{-\epsilon} \,(1-y)^{-\epsilon} \,x_1 \, .
\eea
Here, we have introduced a convenient normalization,
\bea
\sigma_0 = \frac{4\pi \alpha^2}{3Q^2} \frac{\Gamma(1-\epsilon)}{\Gamma(2-2\epsilon)} \left(\frac{3-3\epsilon}{3-2\epsilon}\right) \left(\frac{4\pi\mu^2}{Q^2}\right)^{\epsilon}\, ,
\label{eq:sig-mumu-def}
\eea
which reduces to the inclusive Born cross section for $e^+e^-\to \mu^+\mu^-$ in $D=4$ dimensions.  

As follows from our use of the projection, eq.\ (\ref{eq:spinfierz})  in the computation of partonic fragmentation functions above, we take $\gamma_5$ to be strictly $4$-dimensional to carry out the Dirac trace in computation of $F_\kappa^{(m)}$. This is consistent with Breitenlohner-Maison-'t Hooft-Veltman (BMHV) $\gamma_5$ scheme \cite{tHooft:1972tcz,Breitenlohner:1977hr}. For LO, the heavy quark pair state $\kappa$ can only be octet and we find
\bea
F^{(1)}_{v[8]} =&  \frac{32}{u\bar{u}v\bar{v}}(1-2u)(1-2v) ( 1-\epsilon)\,,\\
F^{(1)}_{a[8]} =&    \frac{32}{u\bar{u}v\bar{v}}\,,\\
F^{(1)}_{\text{eva}A[8]} =&     \frac{-64\epsilon}{u\bar{u}v\bar{v}}\,,\\
F^{(1)}_{\text{eva}B[8]} =&     \frac{32\epsilon}{u\bar{u}v\bar{v}}\,.
\eea
Note that we have contributions from evanescent states, which, although of order $\epsilon$, will be important in subtracting the full long-distance part from NLO partonic cross sections.  The corresponding LO cross sections in eq.\ (\ref{eq:hatsig1}) are
\bea
\frac{d\sigma^{(1)}_{v[8]}}{dx_1} =\, & \sigma_0 \frac{\alpha_s e_Q^2}{Q^2} \pi \frac{32}{u\bar{u}v\bar{v}}(1-2u)(1-2v)(1-\epsilon)\delta(1-x_1)\,,
\label{eq:sig-1-v[8]}\\
\frac{d\sigma^{(1)}_{a[8]}}{dx_1} =\, & \sigma_0 \frac{\alpha_s e_Q^2}{Q^2} \pi \frac{32}{u\bar{u}v\bar{v}}\delta(1-x_1)\,,
\label{eq:sig-1-a[8]}\\
\frac{d\sigma^{(1)}_{\text{eva}A[8]}}{dx_1} =\, &  \sigma_0 \frac{\alpha_s e_Q^2}{Q^2} \pi \frac{-64\epsilon}{u\bar{u}v\bar{v}}\delta(1-x_1)\,,
\label{eq:sig-1-evaA[8]}\\
\frac{d\sigma^{(1)}_{\text{eva}B[8]}}{dx_1} =\, &  \sigma_0 \frac{\alpha_s e_Q^2}{Q^2} \pi \frac{32\epsilon}{u\bar{u}v\bar{v}}\delta(1-x_1) \, .
\label{eq:sig-1-evaB[8]}
\eea
We recall that, by eq.\ (\ref{eq:hat-sgma-1}), the LO cross sections are identical to the LO short-distance coefficients that we use in the computation of the NLO short-distance coefficients (see eq.\ (\ref{eq:order2})).  

Going on to NLO, we compute $F^{(2)}_\kappa$ for $\kappa = v,a$ in color singlet or octet state. Although 
the $F^{(2)}_\kappa$ may be computed for the evanescent states, they will be subleading.  To be specific, to compute non-evanescent $d\hat{\sigma}^{(n)}_\kappa$, we need evanescent contributions to $F^{(m)}_\kappa$ to order $\alpha_s^{n-1}$. To clarify the structure of the $y$ integral, we rewrite $F^{(2)}_\kappa$ for $\kappa = v,a$ in color singlet or octet state as
\bea
\label{eq:Fdecomp}
F^{(2)}_\kappa(x_1,y,u,v,\epsilon) &= \frac{\chi_\kappa (x_1,y,u,v,\epsilon)}{Q^2\,y(1-y)}\,,\nnu
&=\frac{\chi^{(0)}_\kappa (x_1,y,u,v) + \epsilon \,\chi^{(1)}_\kappa (x_1,y,u,v)}{Q^2\,y(1-y)} + \mathcal{O}(\epsilon^2)\,,
\eea
where we have expanded to first order in $\epsilon$ as the higher order terms in $\epsilon$ do not contribute to the NLO cross section in $D=4$. For the term $\epsilon \,\chi^{(1)}$, we note that there are contributions coming from the additional $D-4$ components of the momentum $p_2$, denoted as $\hat{p}_2$. When there is $\gamma_5$ in the $D$-dimensional Dirac trace, such $D-4$ components naturally appear in the BMHV scheme.  As demonstrated in Appendix \ref{appendixPS}, they can simply be replaced by
\bea
\label{eq:p2replace}
\hat{p}_2^2 \to Q^2(1-x_1)y(1-y)\left(\frac{-2\epsilon}{2-2\epsilon}\right)\,.
\eea
To be explicit, $F^{(2)}_\kappa(x_1,y,u,v,\epsilon)$ has $\hat{p}_2^2$ dependent terms with some coefficient $b$ that can be replaced by eq.\ (\ref{eq:p2replace}) as
\bea
F^{(2)}_\kappa(x_1,y,u,v,\epsilon) \supset b(x_1,u,v) \frac{\hat{p}_2^2}{Q^2\,y^2(1-y)^2} &\to \left(\frac{-2\epsilon}{2-2\epsilon}\right)\,b(x_1,u,v)\,(1-x_1) \frac{1}{y(1-y)}\,,\nnu
& = -\frac{\epsilon\,b(x_1,u,v)(1-x_1)}{y(1-y)} + \mathcal{O}(\epsilon^2)\,.
\eea
The term $-b(x_1,u,v)(1-x_1)$ contributes to $\chi_\kappa^{(1)}$ in eq.\ (\ref{eq:Fdecomp}) above.

Using our decomposition of $F^{(2)}_\kappa$  
and also using the symmetry $y \leftrightarrow 1-y$, we can explicitly separate eq.\ (\ref{eq:hatsig2}) into finite and pole pieces as
\bea
\label{eq:sigma-2-NXY}
\frac{d\sigma^{(2)}_\kappa}{dx_1} =& \ \frac{\sigma_0\alpha_s^2 e_Q^2}{Q^2}  \frac{x_1}{4} \,\bigg[-\left(\frac{1}{\epsilon}+\ln\left(\frac{4\pi\mu^2 e^{-\gamma_E}}{Q^2x_1^2(1-x_1)}\right)\right)\chi_\kappa^{(0)}(x_1,1,u,v)\nnu
&\hspace{6cm}+\int dy \,\frac{\chi_\kappa^{(0)}(x_1,y,u,v)}{(1-y)_+} -\chi_\kappa^{(1)}(x_1,1,u,v)\bigg] \, \nnu
\equiv&\ \frac{\sigma_0\alpha_s^2 e_Q^2}{Q^2}\mathcal{N}(x_1,u,v)\left[\left(\frac{1}{\epsilon}+\ln\left(\frac{4\pi\mu^2 e^{-\gamma_E}}{Q^2x_1^2(1-x_1)}\right)\right)X_\kappa(x_1,u,v) + Y_\kappa(x_1,u,v) \right]\,,
\eea
where
\bea
\label{eq:ABcoeff}
\mathcal{N}(x_1,u,v) =& \frac{1}{x_1(1-x_1) \,u\,\bar{u}\,v\,\bar{v}  \, (1-u x_1) \,(1-\bar{u}x_1)\,(1-vx_1)\,(1-\bar{v}x_1)}\, .
\eea
To define the remaining functions on the right in eq.\ (\ref{eq:sigma-2-NXY}), we introduce the notation,
\begin{equation}
u_{1/2}\equiv\left(u-\frac{1}{2}\right)\qquad \text{and} \qquad v_{1/2}\equiv\left(v-\frac{1}{2}\right)\, ,
\end{equation}
in terms of which they are given by
\bea
X_{v[1]}(x_1,u,v)=& -\frac{8}{3} (1-x_1)^2 \,(2-x_1)^2\,,\\
X_{v[8]}(x_1,u,v)=&-\frac{20}{3} (1-x_1)^2 \,(2-x_1)^2-24 x_1^2\left(x_1^4-6 x_1^3+16 x_1^2-20 x_1+10\right) \,u_{1/2} \,v_{1/2} \nnu
&+96 x_1^4\left(x_1^2+3(1-x_1)\right)  \left(u_{1/2}^3 \,v_{1/2}+u_{1/2} \,v_{1/2}^3\right)-384 x_1^6 \,u_{1/2}^3 \,v_{1/2}^3 \,,\\
X_{a[1]}(x_1,u,v)=&-\frac{32}{3}  x_1^2\,(1-x_1)^2 \,u_{1/2} \,v_{1/2} \,,\\
X_{a[8]}(x_1,u,v)=& -6 (2-x_1)^2\left((1-x_1)^4+1\right)-\frac{80}{3}  x_1^2\,(1-x_1)^2\,u_{1/2} \,v_{1/2} \nnu
&+24  x_1^2\,(2-x_1)^2\left(x_1^2+(1-x_1)\right)  \left(u_{1/2}^2+v_{1/2}^2\right)\nnu
&-96 x_1^4\left((1-x_1)^2+1\right)\,u_{1/2}^2 \,v_{1/2}^2   \,,\\
Y_{v[1]}(x_1,u,v)=&(1-x_1)\,(2-x_1)^2\,\left(\frac{2}{3}x_1^2 + 4(1-x_1)\right)  \nnu
&-\frac{8}{3} x_1^2 \,(1-x_1) \,(2-x_1)^2  \left(u_{1/2}^2+v_{1/2}^2\right)+\frac{32}{3} x_1^4 \,(1-x_1) \,u_{1/2}^2\,v_{1/2}^2\,,\\
Y_{v[8]}(x_1,u,v)=&(1-x_1)\,(2-x_1)^2\,\left(\frac{5}{3}x_1^2 + 10(1-x_1)\right) \nnu
&+ 2 x_1^2 \left(x_1^4-36 x_1^3+144 x_1^2-216 x_1+108\right) \,u_{1/2} \,v_{1/2} \nnu
& -\frac{20}{3}x_1^2 \,(1-x_1) \,(2-x_1)^2  \left(u_{1/2}^2+v_{1/2}^2\right) +\frac{80}{3} x_1^4\,(1-x_1) \,u_{1/2}^2 \,v_{1/2}^2\nnu
&-8 x_1^4 \left(x_1^2 + 18 (1-x_1)\right)  \left(u_{1/2}^3 \,v_{1/2}+u_{1/2} \,v_{1/2}^3\right) + 32x_1^6 \,u_{1/2}^3 \,v_{1/2}^3 \,,\\
Y_{a[1]}(x_1,u,v)=&\frac{32}{3}  x_1^2\,(1-x_1)^2 \,u_{1/2} \,v_{1/2}\,,\\
Y_{a[8]}(x_1,u,v)=&-\frac{1}{2}(2-x_1)^2\,\left(11x_1^4-30x_1^3-18x_1^2+96x_1-48\right) + \f{80}{3}x_1^2\,(1-x_1)^2\,u_{1/2} \,v_{1/2}\nnu
&+2x_1^2(2-x_1)^2\left(11x_1^2+12(1-x_1)\right)\left(u_{1/2}^2+v_{1/2}^2\right)\nnu
&-8x_1^4\left(11x_1^2+38(1-x_1)\right)\,u_{1/2}^2 \,v_{1/2}^2\, .
\eea

\subsection{Explicit short-distance coefficients}

We can now present explicit results of short-distance coefficients differential in $x_1$. From eq.\ (\ref{eq:hat-sgma-1}), LO short-distance coefficients differential in $x_1$ are identical to the corresponding partonic cross section, i.e.
\bea
\frac{d\hat{\sigma}^{(1)}_{\kappa}}{dx_1} = \frac{d\sigma^{(1)}_{\kappa}}{dx_1}\,.
\label{eq:hat-sigma-1}
\eea
At NLO, we use eq.\ (\ref{eq:order2}) to derive the $x_1$ differential results,
\bea
\label{eq:hardNLO}
\frac{d\hat{\sigma}^{(2),\overline{\text{MS}}}_{\kappa}}{dx_1} = \frac{d\sigma^{(2)}_{\kappa}}{dx_1}  - \sum_{\zeta = v,a,\text{eva}A,\text{eva}B}\frac{d\hat{\sigma}^{(1)}_{\zeta}}{dx}(x = \frac{x_1}{z},u',v')\otimes_{z;u',v'} \mathcal{D}_{Q\bar{Q}(\zeta) \to Q\bar{Q}(\kappa)}^{(1),\overline{\text{MS}}}(z,u',v',u,v)\,,
\eea 
where we now make the $\overline{\text{MS}}$ scheme dependence explicit.

To compute eq.\ (\ref{eq:hardNLO}) we make use of the results in eqs.\ (\ref{eq:sig-1-v[8]})\ -\ (\ref{eq:sig-1-evaB[8]}) for the Born cross sections at fixed $u$ and $v$, and $\rm \overline{MS}$ partonic fragmentation functions given in eqs.\ (\ref{eq:P-1-evaAv1})\ -\ (\ref{eq:P-1-evaBa8}) and \cite{Kang:2014tta}. We give explicit subtraction terms for $\kappa = v,a$ in color singlet or octet state
\bea
\sum_{\zeta}\frac{d\sigma^{(1)}_{\zeta}}{dx}(x=\frac{x_1}{z},&u',v')\otimes_{z;u',v'} \mathcal{D}_{Q\bar{Q}(\zeta) \to Q\bar{Q}(\kappa)}^{(1),\overline{\text{MS}}} (z,u',v',u,v)\nnu
=& \frac{\sigma_0\alpha_s^2 e_Q^2}{Q^2} \mathcal{N}(x_1,u,v) \left(\left(\f{1}{\varepsilon} + \ln 4\pi e^{-\gamma_E}\right) X_\kappa(x_1,u,v)+ Z_\kappa(x_1,u,v)\right)\,,
\label{eq:sig-1-D-1}
\eea
where $\mathcal{N}(x_1,u,v)$ and $X_\kappa(x_1,u,v)$ are defined above in eq.\ (\ref{eq:ABcoeff}).   The functions $Z_\kappa(x_1,u,v)$ define the extra finite pieces of the subtracton terms, and are given by
\bea
Z_{v[1]}(x_1,u,v) =& \frac{16}{3}(1-x_1)^2(2-x_1)^2\,,\\
Z_{v[8]}(x_1,u,v) =& \frac{40}{3}(1-x_1)^2\,(2-x_1)^2 + 24 x_1^2 \left(x_1^4-6 x_1^3+18 x_1^2-24 x_1+12\right) \,u_{1/2} \,v_{1/2} \nnu
&-96 x_1^4 \left(x_1^2 + 3 (1-x_1)\right)  \left(u_{1/2}^3 \,v_{1/2}+u_{1/2} \,v_{1/2}^3\right) + 384x_1^6 \,u_{1/2}^3 \,v_{1/2}^3 \,,\\
Z_{a[1]}(x_1,u,v) =& \frac{80}{3}  x_1^2\,(1-x_1)^2 \,u_{1/2} \,v_{1/2}\\
Z_{a[8]}(x_1,u,v) =& 6 (1-x_1)^2(2-x_1)^2\left((1-x_1)^2+4\right)+\frac{200}{3}  x_1\,(1-x_1)^2\,u_{1/2} \,v_{1/2} \nnu
&-24  x_1^2\,(1-x_1)^3\,(2-x_1) \left(u_{1/2}^2+v_{1/2}^2\right)+96 x_1^4(1-x_1)^2\,u_{1/2}^2 \,v_{1/2}^2 \,.
\eea
Then, combining the cross section in eq.\ (\ref{eq:sigma-2-NXY}) with its subtraction given by eq.\ (\ref{eq:sig-1-D-1}), according to
 eq.\ (\ref{eq:hardNLO}), our explicit NLO short-distance coefficients for vector or axial states in $\overline{\text{MS}}$ scheme take the form
\bea
\frac{d\hat{\sigma}^{(2),\overline{\text{MS}}}_{\kappa}}{dx_1} =& \frac{\sigma_0\alpha_s^2 e_Q^2}{Q^2}\mathcal{N}(x_1,u,v)\left[\ln\left(\frac{\mu^2}{Q^2x_1^2(1-x_1)}\right)X_\kappa(x_1,u,v) + Y_\kappa(x_1,u,v) -  Z_\kappa(x_1,u,v) \right]\,.
\label{eq:hat-sigma-2-ABC}
\eea
We note that omission of evanescent fragmentation functions from the sum over $\zeta$ in eq.\ (\ref{eq:sig-1-D-1}) would change the values of $Z_\kappa(x_1,u,v)$, but would not change $X_\kappa(x_1,u,v)$ since $\frac{d\sigma^{(1)}_{\zeta}}{dx}$ for evanescent state $\zeta$ is linear in $\epsilon$. Therefore, at this order, evanescent subtractions are not necessary in subtracting the poles. However, they  subtract the finite terms that are sensitive to long-distance dynamics, resulting from a long-distance pole times a term proportional to $\epsilon$ in the hard-scattering function. In fact, at higher orders they would be needed even to subtract the IR poles consistently.    At higher loops, there are terms with the same LO hard part proportional to $\epsilon$, multiplied by multiple poles of higher order evanescent fragmentation functions.

\section{Comparison to NRQCD}
\label{sec:comparison}
The determination of 
short-distance coefficients 
in
 the previous section 
is consistent with any model of factorized long-distance behavior.   As noted in the introduction, fragmentation functions for the heavy quark pair have been computed 
in refs.\ \cite{Ma:2013yla,Ma:2014eja}, assuming the applicability of NRQCD factorization to
 the corresponding matrix elements.    In this section, we shall make use of this assumption, which is certainly valid at NLO.   By combining the fragmentation functions of refs.\ \cite{Ma:2013yla,Ma:2014eja} with the pQCD 
 short-distance coefficients derived in the previous section, eqs.\ (\ref{eq:hat-sigma-1}) and (\ref{eq:hat-sigma-2-ABC}), we
  find cross sections for the relevant NRQCD channels that can be compared directly to cross sections in NRQCD at the $\mathcal{O}(\alpha^2\alpha_s^2)$.  

For two of the channels, ${}^3S_1^{[1]}$ \cite{Cho:1996cg,Keung:1980ev,Yuan:1996ep} and ${}^1S_0^{[8]}$ \cite{Sun:2018yam}, we are lucky enough to have explicit NRQCD calculations at the $\mathcal{O}(\alpha^2\alpha_s^2)$ with which to compare. To this order, these channels begin at NLP, and we will see that the comparison is relatively direct.  We expect and will confirm that the cross sections derived as above, 
by combining pQCD short-distance coefficients and NRQCD fragmentation functions in the ${}^3S_1^{[1]}$ and ${}^1S_0^{[8]}$ channels, have high-energy behaviors that agree precisely with those of the NRQCD calculations.   This equality, while in principle straightforward, requires consistent renormalization procedures along with systematic treatment of the evanescent partonic fragmentation functions introduced in section\ \ref{sec:fact-summary}.
 We begin our discussion by reviewing the relationship between NLP factorization and fixed-order NRQCD calculations.
 
\subsection{Relating NLP factorization to fixed-order NRQCD}
\label{sec:nlp-fo}
Non-relativistic QCD can be applied with or without the presence of a perturbative scale beyond the heavy quark mass.   NRQCD treats the heavy quark mass, $m_Q$, as a hard scale, separating amplitudes and cross sections for these processes into relatively short-distance coefficients, associated with hard scales at and above $\cc{O}(m_Q)$, and universal long-distance matrix elements (LDMEs) associated with soft scales at and below $\cc{O}(m_Q v)$. Schematically, the cross section can be factorized at NRQCD factorization scale $\mu_\Lambda \sim m_Q$ as
\bea
\label{eq:NRQCDfact}
\sigma^{H}_{\text{NRQCD}} = \sum_\nu \hat{f}_{Q\bar{Q}(\nu)}(m_Q,\mu_{\Lambda}) \langle \mathcal{O}_{Q\bar{Q}(\nu)}^H(\mu_{\Lambda})\rangle\, ,
\eea
where $\hat{f}_{Q\bar{Q}(\nu)}$ is a perturbative coefficient describing the production of a heavy quark pair in NRQCD state $\nu$, and $\langle \mathcal{O}\rangle$ is the LDME describing the formation of the observed heavy quarkonium $H$ from the heavy quark pair state $\nu$. These LDMEs are scaled in powers of the heavy quark pair's relative velocity $v \ll 1$. Applications of NRQCD have resolved tensions between theoretical predictions and experimental measurements, using a limited number of LDMEs \cite{Brambilla:2010cs,Brambilla:2004wf}. Puzzles remain, however, especially involving polarization and the tension between LDME values required to fit different production processes \cite{Brambilla:2010cs,Butenschoen:2011yh,Butenschoen:2012qr}.     Since these tensions often involve some larger perturbative scale, $E_H$, where heavy quark pair fragmentation is important \cite{Ma:2014svb}, it is natural to reconsider these cross sections in the extended pQCD formalism discussed above.

We can carry out matching for our pQCD fragmentation functions in eq.\ (\ref{eq:calD-def}) in terms of the same universal LDMEs of NRQCD at the input scale $\mu_0 \sim 2m_Q$,  \cite{Kang:2014tta, Kang:2014pya,Ma:2013yla,Ma:2014eja} as
\bea
\label{eq:DQQNR}
\mathcal{D}_{Q\bar{Q}(\kappa)\to H}(z,u,v;m_Q,\mu_0)\ =\ \sum_{\nu} \hat{d}_{Q\bar{Q}(\kappa) \to Q\bar{Q}(\nu)}(z,u,v;m_Q,\mu_0,\mu_\Lambda) \langle \mathcal{O}_{Q\bar{Q}(\nu)}^H (\mu_{\Lambda})\rangle\,,
\eea
where $\kappa$ and $\nu$ label a relativistic pQCD state and a non-relativistic NRQCD state, respectively, also appearing in eq.~(\ref{eq:pqcd_fac}) and in eq.~(\ref{eq:NRQCDfact}). Here, the matching coefficients $\hat{d}_{Q\bar{Q}(\kappa) \to Q\bar{Q}(\nu)}(z,u,v;m_Q,\mu_0,\mu_\Lambda)$ are perturbatively computable.
This matching is analogous to the NRQCD treatment of single-parton fragmentation, as developed in refs.\ \cite{Braaten:1996rp,Bodwin:2014gia},
\bea
\label{DNR}
D_{i \to J/\psi}(z;m_Q,\mu_0)\ =\ \sum_\nu \hat{d}_{i \to Q\bar{Q}(\nu)}(z;m_Q,\mu_0,\mu_\Lambda) \langle \mathcal{O}_{Q\bar{Q}(\nu)}^H(\mu_\Lambda)\rangle\,,
\eea
for $i=q,\bar{q},g$.
Using these models for fragmentation functions 
at $\mu=\mu_0$, we can express the single-particle  inclusive cross section in the notation of eq.\ (\ref{eq:QCDfact}) at any fixed order\footnote{Once again, we emphasize that the true motivation for NLP pQCD factorization comes from resummation of large logarithms of $\ln \frac{m_Q^2}{E_H^2}$. The fixed order expressions given here at $\mu=\mu_0$ enable us to check NLP factorization.  } 
in terms of a limited number of non-perturbative NRQCD LDMEs as
\bea
\label{eq:QCDLDME}
\sigma^{H}_{\text{pQCD}} =& \sum_{i=q,\bar{q},g} \sum_{\nu} \hat{\sigma}_{i}(\mu_0) \otimes_z \hat{d}_{i \to Q\bar{Q}(\nu)}(\mu_0) \langle \mathcal{O}_{Q\bar{Q}(\nu)}^H\rangle 
\nnu
& \hspace{25mm}+ \sum_\kappa \sum_\nu \hat{\sigma}_{Q\bar{Q}(\kappa)}(\mu_0) \otimes_{z;u,v} \hat{d}_{Q\bar{Q}(\kappa) \to Q\bar{Q}(\nu)}(\mu_0) \langle \mathcal{O}_{Q\bar{Q}(\nu)}^H \rangle \nnu
=&\sum_\nu \left(\sum_{i=q,\bar{q},g} \hat{\sigma}_{i}(\mu_0) \otimes_z \hat{d}_{i \to Q\bar{Q}(\nu)}(\mu_0) + \sum_\kappa \hat{\sigma}_{Q\bar{Q}(\kappa)} (\mu_0)\otimes_{z;u,v} \hat{d}_{Q\bar{Q}(\kappa) \to Q\bar{Q}(\nu)} (\mu_0)\right) \langle \mathcal{O}_{Q\bar{Q}(\nu)}^H \rangle\,,
\eea
where we suppress dependence on variables other than the factorization scale $\mu = \mu_0$.

Expanding in relative velocities up to $v^4$, the full set of LDMEs
associated with heavy quark pair states is
\bea
\label{channels}
{}^3S_1^{[1]},{}^3S_1^{[8]},{}^1S_0^{[8]},{}^3P_J^{[8]}\, .
\eea
This set provides predictions in terms of only a few NRQCD parameters.  As emphasized in refs.\ \cite{Ma:2013yla,Ma:2014eja}, computing fragmentation functions using NRQCD in principle
 enables us to go from fitting several non-perturbative LP and NLP fragmentation functions to determining a small
 number of LDMEs.   Our discussion here checks the consistency of this procedure.

Comparing the basic NRQCD relation, eq.\ (\ref{eq:NRQCDfact}) with the high-$E_H$ factorization with NRQCD input for the fragmentation functions, eq.\ (\ref{eq:QCDLDME}), one can expect an order-by-order relation between NRQCD and NLP factorization short-distance coefficient functions,
\bea
\label{eq:QCDNRQCD}
 \hat{f}_{Q\bar{Q}(\nu)}(\mu_\Lambda) &=\sum_{i=q,\bar{q},g} \hat{\sigma}_{i} (\mu_0)\otimes_z \hat{d}_{i \to Q\bar{Q}(\nu)} (\mu_0,\mu_\Lambda)
 \nnu
 & \hspace{5mm} + \sum_\kappa \hat{\sigma}_{Q\bar{Q}(\kappa)} (\mu_0)\otimes_{z;u,v} \hat{d}_{Q\bar{Q}(\kappa) \to Q\bar{Q}(\nu)}(\mu_0,\mu_\Lambda) + \mathcal{O}\left(\frac{m_Q^2}{E_H^2}\right) \, .
 \eea
Clearly, the full NRQCD calculation contains more information than the NLP factorized cross section at the fixed order in $\alpha_s$.   All such information, however, appears  beyond NLP.  This was confirmed numerically for high-$p_T$ production at hadronic colliders in selected channels by ref.\ \cite{Ma:2014svb}, using LO pQCD short-distance coefficients.  Such comparisons serve both as a test of the NLP formalism, and as a tool for studying the approach to high-energy behavior. In the remainder of this section, we use the results of section \ref{sec:lo-nlo} to analytically confirm eq.\ (\ref{eq:QCDNRQCD})  for the channels  ${}^3S_1^{[1]}$ and ${}^1S_0^{[8]}$ .

\subsection{Input fragmentation functions and factorized cross sections}
\label{subsec:inputFF}

For a given NRQCD channel, we compute the fixed-order pQCD prediction of the $x_H = 2E_H/Q$ differential version of eq.\ (\ref{eq:QCDLDME}) to $\mathcal{O}(\alpha^2\alpha_s^2)$. To do so, we need matching coefficients for fragmentation functions, $\hat{d}$, in eq.\ (\ref{DNR}) to $\mathcal{O}(\alpha_s)$. As our gluon-associated process will only involve NLP scaling to the order we consider, we only need to consider NRQCD factorization of the pair fragmentation functions.

It will be convenient below to factor out mass and coupling dependence compared to the $\hat{d}$ in eq.\ (\ref{eq:DQQNR}), defining perturbative coefficients $\hat{d}^{(i)}$ by
\bea
\label{inputFF_dec}
\cc{D}_{[Q\bar{Q}(\kappa)]\to H }(z,u,v;m_Q,\mu_0)=&\sum_\nu \bigg(\hat{d}^{(0)}_{[Q\bar{Q}(\kappa)]\to [Q\bar{Q}(\nu)]} (z,u,v; m_Q, \mu_0,\mu_\Lambda) \nnu
&\hspace{-35mm}+ \left(\f{\as}{\pi}\bigg)\hat{d}^{(1),\overline{\text{MS}}}_{[Q\bar{Q}(\kappa)]\to [Q\bar{Q}(\nu)]}(z,u,v;m_Q, \mu_0,\mu_\Lambda) + \mathcal{O}(\alpha_s^2)\right)\times\f{\langle \cc{O}^{H}_{[Q\bar{Q}(\nu)]}(\mu_\Lambda)\rangle}{m_Q^{2L+1}}\,.
\eea
Note that the input fragmentation function does not have $u'$ and $v'$ dependence that is present in the partonic fragmentation functions in eq.\ (\ref{eq:partDMS}). This is because the final state in
 the input fragmentation function is a heavy quarkonium. Of course, unlike the partonic fragmentation function, the input fragmentation function is also a non-perturbative object with perturbative matching coefficient extracted. 

We now apply the notation of eq.\ (\ref{inputFF_dec}) to the general cross section eq.\ (\ref{eq:QCDLDME})  for heavy quarkonium production at fixed $x_H$. We can then isolate the contribution from a fixed intermediate NRQCD state $\nu$ to a heavy quarkonium production cross section, at fixed $x_H$  at order $\mathcal{O}(\alpha^2\alpha_s^2)$, as
\bea
\label{eq:xHpQCD}
\f{d\sigma^{(2),\text{pQCD}}_{e^+e^- \to[ Q\bar{Q}(\nu)]\to Hgg}}{dx_H} &= \bigg[\sum_\kappa\f{d\hat{\sigma}^{(1)}_{\kappa}}{dx_1} (x_1 = \f{x_H}{z},u,v,\mu_0) \otimes_{z;u,v} \left(\frac{\alpha_s}{\pi}\right)\hat{d}^{(1),\overline{\text{MS}}}_{[Q\bar{Q}(\kappa)]\to [Q\bar{Q}(\nu)]}(z,u,v; m_Q,\mu_0, \mu_\Lambda)\nnu
+\sum_\kappa\f{d\hat{\sigma}^{(2),\overline{\text{MS}}}_{\kappa}}{dx_1}&(x_1 = \f{x_H}{z},u,v,\mu_0) \otimes_{z;u,v}\hat{d}^{(0),\overline{\text{MS}}}_{[Q\bar{Q}(\kappa)]\to [Q\bar{Q}(\nu)]}(z,u,v ; m_Q, \mu_0,\mu_\Lambda)\bigg]\f{\langle \cc{O}^{H}_{[Q\bar{Q}(\nu)]}(\mu_\Lambda)\rangle}{m_Q^{2L+1}}\nnu
& \equiv\  \sigma_0 \f{\alpha_s^2e_Q^2}{Q^{2}}\,  \left (
\hat{k}_\nu^{(1)}(x_H;\mu_0,\mu_\Lambda)+\hat{k}_\nu^{(2)}(x_H;\mu_0,\mu_\Lambda)\right)\f{\langle \cc{O}^{H}_{[Q\bar{Q}(\nu)]}(\mu_\Lambda)\rangle}{m_Q^{2L+1}} \,.
\eea
The second relation defines the functions $\hat{k}_\nu^{(i)}(x_H;\mu_0,\mu_\Lambda)$, which result from the convolutions of $\hat\sigma^{(1)}$ with $\hat{d}^{(1)}$ and $\hat\sigma^{(2)}$ with $\hat{d}^{(0)}$, respectively.   Their normalization is set by separating the overall factor, $\sigma_0$ $\frac{\alpha_s^2 e_Q^2}{Q^2} \f{\langle\mathcal{O}\rangle}{m_Q^{2L+1}}$, with $\sigma_0$
given in eq.\ (\ref{eq:sig-mumu-def}) above.

To facilitate the comparison of these results with direct NRQCD calculations, we rescale NRQCD coefficients,  $\hat{f}$ in eq.\ (\ref{eq:NRQCDfact}), with the same overall normalization as in eq.\ (\ref{eq:xHpQCD}),
\bea
\f{d\sigma^{(2),\text{NRQCD}}_{e^+e^- \to Q\bar{Q}(\nu)gg\to H gg}}{dx_H} =& \sigma_0 \f{\alpha_s^2e_Q^2}{Q^2} \, \hat{f}_\nu(x_H,m_Q,\mu_\Lambda)\,\,\f{\langle \cc{O}^{H}_{[Q\bar{Q}(\nu)]}(\mu_\Lambda)\rangle}{m_Q^{2L+1}}\,.
\eea
We conclude that for each channel $\nu$,
\bea
\label{eq:approx}
\hat{f}_{\nu}(x_H,m_Q,\mu_\Lambda) = \hat{k}^{(1)}_{\nu}(x_H;\mu_0,\mu_\Lambda) + \hat{k}^{(2)}_{\nu}(x_H;\mu_0,\mu_\Lambda)  + \mathcal{O}(r)\,.
\eea
This is the $\mathcal{O}(\alpha_s^2)$ version of eq.\ (\ref{eq:QCDNRQCD}). Here, the measure of higher-power corrections to the large-$E_H$ behavior is given by
\bea
r \ =\ \frac{m_Q^2}{E_H^2}\,.
\label{eq:r-def}
\eea
We will test eq.\ (\ref{eq:approx}) below for channels in which an explicit NRQCD calculation is available.

The first step in verifying eq.\ (\ref{eq:approx}) is to recall explicit results for $\hat{d}^{(0)}$ and $\hat{d}^{(1)}$.   These can be found in the appendices of \cite{Ma:2013yla,Ma:2014eja}\footnote{The results in \cite{Ma:2013yla,Ma:2014eja} are given in terms of $\zeta_1 = 2u - 1$ and $\zeta_2 = 2v - 1$.}. There, however, the $\hat{d}^{(1)}$ were presented using the Kreimer $\gamma_5$ scheme of refs.\ \cite{Kreimer:1989ke,Korner:1991sx,Kreimer:1993bh} in $\overline{\text{MS}}$ subtraction scheme. To be consistent with our factorization procedure for the pQCD short-distance coefficients above, we must recompute these coefficients in the BMHV $\gamma_5$ scheme in $\overline{\text{MS}}$ subtraction scheme. To be self-contained, we present all the 
functions  $\hat{d}$ used in this paper, some of which are different from \cite{Ma:2013yla,Ma:2014eja} due to the difference in $\gamma_5$ scheme. We present the ones needed
 for ${}^3S_1^{[1]}$ and ${}^1S_0^{[8]}$ here, but list all the other relevant ones in Appendix \ref{appendixhatd}. Suppressing the arguments, they are given as
\bea
\hat{d}^{(0)}_{[Q\bar{Q}(v[1])]\to [Q\bar{Q}({}^3S_1^{[1]})]} =& \frac{1}{24}\delta(u-\frac{1}{2})\delta(v-\frac{1}{2})\delta(1-z)\,,\\
\hat{d}^{(0)}_{[Q\bar{Q}(a[8])]\to [Q\bar{Q}({}^1S_0^{[8]})]} =& \frac{1}{64}\delta(u-\frac{1}{2})\delta(v-\frac{1}{2})\delta(1-z)\,,\\
\hat{d}^{(1),\overline{\text{MS}}}_{[Q\bar{Q}(v[8])]\to [Q\bar{Q}({}^3S_1^{[1]})], z\neq 1} =& \frac{C_F}{24}\frac{1}{N_c^2-1}\Delta_{-}^{[1]}\frac{z}{(1-z)}\left(\ln\frac{\mu_0^2}{4m_Q^2(1-z)^2}+2z^2-4z+1\right)\,,\\
\hat{d}^{(1),\overline{\text{MS}}}_{[Q\bar{Q}(v[8])]\to [Q\bar{Q}({}^1S_0^{[8]})], z\neq 1} =& \frac{C_F}{8}\frac{1}{(N_c^2-1)^2}\Delta_+^{[8]}z(1-z)\ln\frac{\mu_0^2}{4m_Q^2(1-z)^2}\,,\\
\hat{d}^{(1),\overline{\text{MS}}}_{[Q\bar{Q}(a[8])]\to [Q\bar{Q}({}^3S_1^{[1]})], z\neq 1}  =& \frac{C_F}{24}\frac{1}{N_c^2-1}\Delta_{+}^{[1]}z(1-z)\left(\ln\frac{\mu_0^2}{4m_Q^2(1-z)^2}+2\right) \,,\\
\hat{d}^{(1),\overline{\text{MS}}}_{[Q\bar{Q}(a[8])]\to [Q\bar{Q}({}^1S_0^{[8]})], z\neq 1} =& \frac{C_F}{8}\frac{1}{(N_c^2-1)^2}\Delta_-^{[8]}\frac{z}{1-z}\left(\ln\frac{\mu_0^2}{4m_Q^2(1-z)^2}-1\right)\,,
\eea
where
\bea
\Delta_\pm^{[1]} =&\,  [\delta(u-\frac{z}{2} ) \pm \delta(\bar{u}-\frac{z}{2})][\delta(v-\frac{z}{2}) \pm \delta(\bar{v}-\frac{z}{2})]\,,\\
\Delta_\pm^{[8]} =&\,  (N_c^2-2)[\delta(u-\frac{z}{2})\delta(v-\frac{z}{2}) + \delta(\bar{u}-\frac{z}{2})\delta(\bar{v}-\frac{z}{2})] \nnu
&\hspace{4cm}\mp 2[\delta(u-\frac{z}{2})\delta(\bar{v}-\frac{z}{2}) + \delta(\bar{u}-\frac{z}{2})\delta(v-\frac{z}{2})]\,,\\
\Delta_\pm^{[8]'} =& -\frac{z}{2}\bigg\{(N_c^2-2)\bigg[\delta'(u-\frac{z}{2})\delta(v-\frac{z}{2}) +\delta(u-\frac{z}{2})\delta'(v-\frac{z}{2}) \\
&\hspace{4cm}+ \delta'(\bar{u}-\frac{z}{2})\delta(\bar{v}-\frac{z}{2})+\delta(\bar{u}-\frac{z}{2})\delta'(\bar{v}-\frac{z}{2})\bigg] \nnu
&\mp 2\bigg[\delta'(u-\frac{z}{2})\delta(\bar{v}-\frac{z}{2}) +\delta(u-\frac{z}{2})\delta'(\bar{v}-\frac{z}{2}) \nnu
&\hspace{4cm}+ \delta'(\bar{u}-\frac{z}{2})\delta(v-\frac{z}{2})+\delta(\bar{u}-\frac{z}{2})\delta'(v-\frac{z}{2})\bigg]\bigg\}\,,\\
\Delta_\mp^{[8]''} =& \, \frac{z^2}{4} \bigg\{(N_c^2-2)\bigg[\delta'(u-\frac{z}{2})\delta'(v-\frac{z}{2}) + \delta'(\bar{u}-\frac{z}{2})\delta'(\bar{v}-\frac{z}{2})\bigg] \nnu
&\hspace{4cm}\mp 2\bigg[\delta'(u-\frac{z}{2})\delta'(\bar{v}-\frac{z}{2}) + \delta'(\bar{u}-\frac{z}{2})\delta'(v-\frac{z}{2})\bigg]\bigg\}\,.
\eea
Note that $\Delta_{\pm}^{[1,8]}$ are identical to 
those
found in eqs.\ (\ref{eq:Dpm1}) and\ (\ref{eq:Dpm8}) with $u'=v'=1/2$. Note also that we have dropped $\delta(1-z)$ dependent terms in these NLO matching coefficients.  This is because the LO cross sections $d\hat \sigma^{(1)}/dx_1$ are all proportional to $\delta(1-x_1)$.  
Then in eq.\ (\ref{eq:xHpQCD}), any $\delta(1-z)$ terms in $\hat{d}^{(1)}$ contribute only at $x_H=1$, and thus are not included in this study.

We also do not need to compute input fragmentation functions for evanescent intermediate states. We recall that the LO short-distance coefficient functions given in eqs.\ (\ref{eq:sig-1-evaA[8]}) and (\ref{eq:sig-1-evaB[8]}) for evanescent states are proportional to $\epsilon$.   These terms contribute to the partonic cross section at order $\epsilon^0$ because they multiply the infrared pole of the evanescent partonic fragmentation functions.   All input fragmentation functions calculated from NRQCD, however, evanescent or four-dimensional, are finite after renormalization, and the poles of the partonic calculation are replaced by finite logarithms.   The corresponding terms thus remain of order $\epsilon$, and vanish in four dimensions.  
 
Convolving the short-distance coefficients and the $\hat{d}$ presented here, we can write explicit expressions for the $\hat{k}_\nu^{(i)}(x_H;\mu_0,\mu_\Lambda)$ in eq.\ (\ref{eq:xHpQCD}) for the relevant NRQCD channels. We find for ${}^3S_1^{[1]}$ and ${}^1S_0^{[8]}$,
\bea
\label{eq:k13S1}
\hat{k}^{(1)}_{{}^3S_1^{[1]}}(x_H;\mu_0,\mu_\Lambda) =& \frac{256}{9} \f{(1-x_H)}{x_H(2-x_H)^2}\left(  \ln \left( \f{\mu_0^2}{4m_Q^2(1-x_H)^2} \right) +x_H^2-2x_H+\frac{3}{2}\right)\,,\\
\label{eq:k23S1}
\hat{k}^{(2)}_{{}^3S_1^{[1]}}(x_H;\mu_0,\mu_\Lambda) =& -\frac{256}{9} \f{(1-x_H)}{x_H(2-x_H)^2}\left(  \ln \left( \f{\mu_0^2}{E_H^24(1-x_H) } \right) -\frac{x_H^2+2x_H-2}{4(1-x_H)}\right)\,,\\
\label{eq:k11S0}
\hat{k}^{(1)}_{{}^1S_0^{[8]}}(x_H;\mu_0,\mu_\Lambda) =&\f{2}{x_H(1-x_H)(2-x_H)^2}\bigg(12\left((1-x_H)^4+1\right)\ln \left( \f{\mu_0^2}{4m_Q^2(1-x_H)^2} \right) -12\bigg)\,,\\
\label{eq:k21S0}
\hat{k}^{(2)}_{{}^1S_0^{[8]}}(x_H;\mu_0,\mu_\Lambda) =&- \f{2}{x_H(1-x_H)(2-x_H)^2}\bigg(  12\left((1-x_H)^4+1\right)\ln \left( \f{\mu_0^2}{E_H^24(1-x_H) } \right) \nnu
&\hspace{3cm}+ 23x_H^4-78x_H^3+102x_H^2-48x_H+12\bigg)\,.
\eea
The results for other NRQCD channels are included  in Appendix \ref{appendixkterms}. 

As noted above on the right-hand side of eq.\ (\ref{eq:approx}), the scale $\mu_0$ in the functions $\hat{k}$ is the pQCD factorization scale, which is not present in the NRQCD calculation on the left-hand side. Therefore, we expect that the sum of the two terms on the right-hand side is also independent of $\mu_0$. We indeed find that the $\mu_0$-dependence of $k_\nu^{(1)}(x_H;\mu_0,\mu_\Lambda)$ and  $k_\nu^{(2)}(x_H;\mu_0,\mu_\Lambda)$ cancel. 
Although we factorized our pair fragmentation functions using NRQCD at the input scale $\mu_0 \sim 2m_Q$, the $\mu_0$-independence allows us to choose any value for our fixed order expression.    This would be relevant at higher orders, where taking an appropriate choice of $\mu_0$, we can control whether large logarithms appear in the short-distance coefficients or the fragmentation functions.
In our case, we find that 
\bea
\hat{k}^{(2)}\propto d\hat{\sigma}^{(2)} \otimes_{z;u,v} \mathcal{D}^{(0)} &\supset \ln \frac{\mu_0^2}{E_H^2(1-x_H)}\,,\\
\hat{k}^{(1)}\propto d\hat{\sigma}^{(1)} \otimes_{z;u,v} \mathcal{D}^{(1)} &\supset -\, \ln \frac{\mu_0^2}{4m_Q^2(1-x_H)^2}\,, 
\eea
and we take $\mu_0 = E_H$ to remove large logarithms from the short-distance coefficients. Notice that there are additional logarithms coming from the threshold limit $x_H\to 1$. Meaningful comparison to the data near such endpoints can only be made after resumming these large logarithms, which can be done in principle by combining threshold resummation techniques to our work \cite{Beneke:1997qw,Fleming:2003gt,Fleming:2006cd,Leibovich:2007vr,Ma:2017xno}. Aside from such threshold logarithms, now that large logarithms only remain in the fragmentation functions, one can try to evolve them to $E_H$ and resum the large logarithms of $\ln (E_H/2m_Q)$. In this way, QCD factorization demonstrates how the natural scale choice appears. 

Of course, both NRQCD and QCD calculations also have a renormalization scale, $\mu_r$. In NRQCD calculations 
of gluon-associated $e^+e^-$ processes,
the renormalization scale is chosen to be $\mu_r = 2m_Q$ or $\mu_r = Q/2$.    For example, refs.\ \cite{Ma:2008gq,Zhang:2009ym,Gong:2009kp} calculate  the total cross section, integrated over energy $E_H$, and are thus left with only $m_Q$ and $Q$ as the relevant scales.  For our energy fraction distribution, we will identify the renormalization scale with the factorization scale, $\mu_r = \mu_0 = E_H$, for both NLP predictions and NRQCD calculations.

\subsection{Comparisons for ${}^3 S_1^{[1]}$ and ${}^1 S_0^{[8]}$ channels}

We are now ready to compare to the NRQCD results in closed form for the channels ${}^3 S_1^{[1]}$ and ${}^1 S_0^{[8]}$ at $\mathcal{O}(\alpha^2\alpha_s^2)$. 

From \cite{Cho:1996cg,Keung:1980ev,Yuan:1996ep} for ${}^3S_1^{[1]}$ and more recently from \cite{Sun:2018yam} for ${}^1S_0^{[8]}$, we then find for $x_H < 1$,
\bea
\label{eq:nrqcdpred}
\hat{f}_{{}^3 S_1^{[1]}}(x_H,m_Q,\mu_\Lambda)|_{x_H\neq 1} =&  \f{64}{9} \f{1}{(2r x_H^2-x_H)^3(2-x_H)^2}\Bigg\{(2r x_H^2-x_H)x_H\sqrt{1-4r}\nnu
&\times\bigg[4+4r x_H^2(5+r x_H^2(7+4r x_H^2)) - 12(1+r x_H^2)(1+2r x_H^2)x_H \nnu
&+ (13+14r x_H^2)x_H^2 - 4x_H^3\bigg] -\ln\left(\f{r x_H^2+x_+-1}{r x_H^2+x_--1}\right) 4(1+r x_H^2-x_H)\nnu
&\times\bigg[2(r x_H^2-1)r x_H^2(1+4r x_H^2(2+r x_H^2))\nnu
&-2r x_H^2(-5+2r x_H^2+6r^2x_H^4)x_H - (1+r x_H^2(1-5r x_H^2))x_H^2\bigg]\Bigg\} \,,\\
\label{eq:nrqcdpred2}
\hat{f}_{{}^1 S_0^{[8]}}(x_H,m_Q,\mu_\Lambda)|_{x_H\neq 1} =&  \frac{2}{(1-r x_H^2)^3(1+r x_H^2-x_H)}\bigg\{12(1-r x_H^2)^3 \ln \frac{x_+^2(1-x_-)}{(1-x_+)(1+r x_H^2-x_H)} \nnu
 &+x_H\sqrt{1-4r}(x_H^2-6x_H-24r^2 x_H^4+50r x_H^2-18)\bigg\}\nnu
&+\frac{2}{(1-r x_H^2)^3(x_H-2r x_H^2)^3(2-x_H)^2}\bigg\{\sqrt{x_H^2-4r x_H^2}\nnu
&\bigg[ 16 rx_H^2 (3 - 9 r x_H^2 + 9 r^2x_H^4 + 24 r^3x_H^6 - 28 r^4x_H^8 + 9 r^5x_H^{10})
\nnu
& - 8 (3 - 3 r x_H^2 - 9 r^2 x_H^4 + 120 r^3 x_H^6- 94 r^4 x_H^8+ 27 r^5 x_H^{10}) x_H \nnu
&+ 4 (6 - 15 r x_H^2 + 162 r^2x_H^4 - 75 r^3x_H^6 + 22 r^4x_H^8) x_H^2
\nnu
&- 2 (3 + 90 r x_H^2 + 23 r^2x_H^4 + 4 r^3x_H^6) x_H^3 +
 2 (12 + 25 rx_H^2 + 3 r^2x_H^4) x_H^4 \nnu
 &- (9 + 5 rx_H^2) x_H^5 + x_H^6\bigg]
\nnu
&+ 12(1-r x_H^2)^2\ln{\frac{r x_H^2+x_+-1}{r x_H^2+x_--1}}\,\nnu
&\times\bigg[ 4 r x_H^2 (1 - 2 r x_H^2 - 6 r^2x_H^4 + 2 r^3x_H^6 - 3 r^4 x_H^8) \nnu
&+ 8 r^2x_H^4 (6 + rx_H^2 + 3 r^2x_H^4) x_H
- 2 (1 + 12 r x_H^2 + 15 r^2x_H^4 + 12 r^3x_H^6) x_H^2 \nnu
&+ 2 (3 + 9 rx_H^2 + 8 r^2x_H^4) x_H^3 -
 2 (2 + 3 rx_H^2) x_H^4 + x_H^5 \bigg]\bigg\}\,,
\eea
where $r$ is defined in eq.\ (\ref{eq:r-def}) and
\bea
 x_\pm = \f{1}{2}(2-x_H\pm x_H\sqrt{1 - 4r})\,.
\eea
With finite heavy quark mass, $x_H$ has a range 
\bea
4m_Q/Q < x_H < 1+4m_Q^2/Q^2\, .
\eea
Expanding these results in $r$, we compare them to the corresponding sums of $\hat{k}^{(1)}$ and $\hat{k}^{(2)}$ in eqs.\ (\ref{eq:k13S1}) to \ (\ref{eq:k21S0}) away from $x_H = 1$.   The results are, as expected, consistent with eq.\ (\ref{eq:approx}),
\bea
\label{eq:nrqcdmless1}
\hat{f}_{{}^3S_1^{[1]}}(x_H,m_Q,\mu_\Lambda)|_{x_H\neq 1}  =& \f{64}{9}\bigg[ \f{4(1-x_H)}{x_H(2-x_H)^2}\ln\left(\f{x_H^2}{(1-x_H)r}\right) -\f{4x_H^3-13x_H^2 +12x_H-4}{x_H(2-x_H)^2}\bigg] + \mathcal{O}(r)\nnu
=& \hat{k}_{{}^3S_1^{[1]}}^{(1)}(x_H,r,\mu)+\hat{k}_{{}^3S_1^{[1]}}^{(2)}(x_H,r,\mu)+ \mathcal{O}(r) \,,\\
\label{eq:nrqcdmless2}
\hat{f}_{{}^1S_0^{[8]}}(x_H,m_Q,\mu_\Lambda)|_{x_H\neq 1}  =& \f{2}{x_H(1-x_H)(2-x_H)^2}\bigg[ 12\left((1-x_H)^4+1\right)\ln\left(\f{x_H^2}{(1-x_H)r}\right) \nnu
&- 23x_H^4+78x_H^3-102x_H^2 +48x_H-24\bigg] + \mathcal{O}(r)\nnu
=& \hat{k}_{{}^1S_0^{[8]}}^{(1)}(x_H,r,\mu)+\hat{k}_{{}^1S_0^{[8]}}^{(2)}(x_H,r,\mu)+ \mathcal{O}(r) \,.
\eea
In summary, for the cases where it can be checked, the pQCD factorization formalism successfully reproduces the correct, and reasonably non-trivial, high-energy behavior of the full calculation.   The results of Appendix \ref{appendixkterms} can also be used in eq.\ (\ref{eq:xHpQCD}) to give new closed expressions for the high energy behavior of the channels for which explicit calculations do not exist.

\section{Numerical results}
\label{sec:numerical}
In the following, we carry out a few numerical investigations of the results of the preceding sections. We begin by studying the $x_H$ distribution in the ${}^3S_1^{[1]}$ channel, to illustrate the approach of the fixed order pQCD cross section to the full NRQCD result as $E_H$ increases relative to $m_H$.  In section\ \ref{subsec:dist-sH}, we study energy fraction $x_H$ distributions of different NRQCD channels for $H=J/\psi$, and compare to the $x_H$ distribution of Belle data for $J/\psi$ production at $Q = 10.6$ GeV. In section\ \ref{sec:doublelog}, we explore the significance of the logarithmic corrections associated with the evolution of the heavy quark pair fragmentation functions.

\subsection{Approach to full NRQCD result}
\label{sub:approach}
\bef
\includegraphics[width=6in]{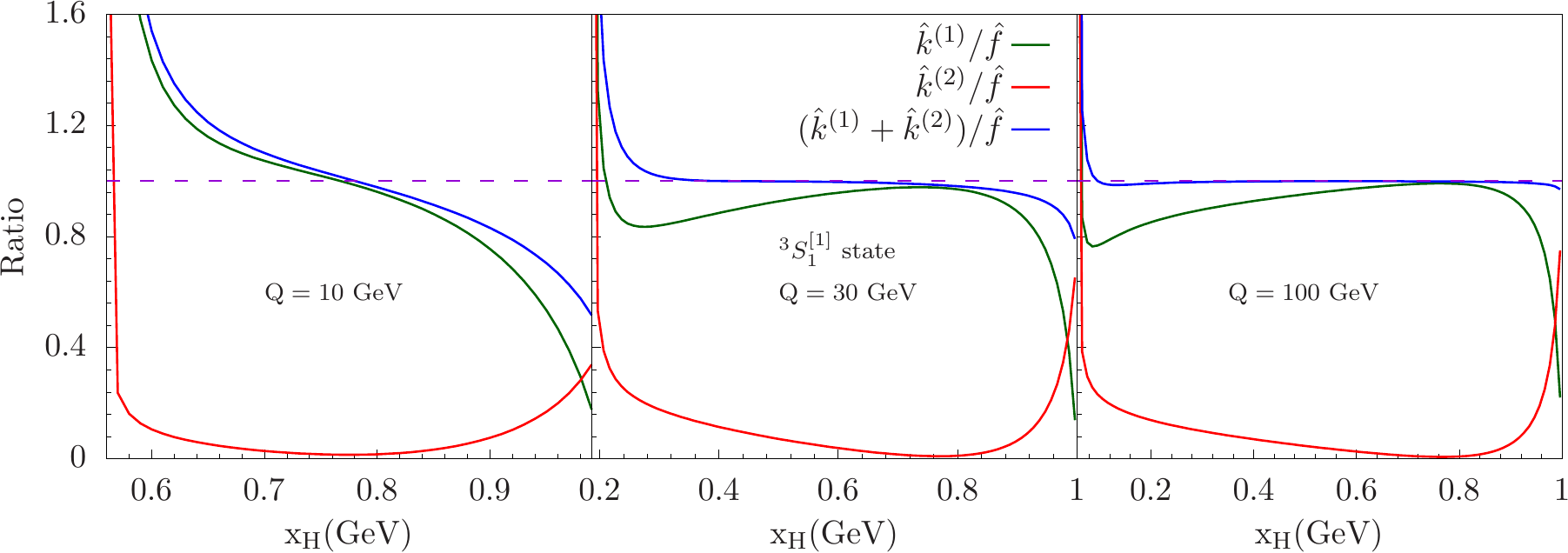}
\caption{The ratios of pQCD and NRQCD results, $\hat{k}^{(i)}/\hat{f}$ for ${}^3S_1^{[1]}$, from eqs.\ (\ref{eq:k13S1}), (\ref{eq:k23S1}) and (\ref{eq:nrqcdpred}), at CM energies $10$, $30$, and $100$ GeV.}
\label{Fig:compare}
\eef

Figure \ref{Fig:compare} shows the ratios of the pQCD functions, $\hat{k}_{{}^3S_1^{[1]}}^{(1)}$ and $\hat{k}_{{}^3S_1^{[1]}}^{(2)}$ given in eqs. \ (\ref{eq:k13S1}) and \ (\ref{eq:k23S1}), and their sum, to the corresponding full NRQCD function $\hat{f}_{{}^3 S_1^{[1]}}$ given in eq.\ (\ref{eq:nrqcdpred}). Here, we set the charm quark mass $m_Q = m_c = 1.4$ GeV and we plot the ratio over the full physical region, $4m_Q/Q < x_H < 1+4m_Q^2/Q^2$, for  three representative values of the CM energy, $Q=\sqrt{q^2}\ =\ 10,\, 30$ and $100$ GeV.  

The divergence of the ratio in the small-$x_H$ limit is due to the vanishing of $\hat{f}_{{}^3S_1^{[1]}}$ in eq.\ (\ref{eq:nrqcdpred}) for $x_{H} \to 4m_Q/Q $. The cross section no longer vanishes when power corrections in $\cc{O}(\frac{m_{H}^2}{E_{H}^2})$ are dropped, and thus the ratio produces a divergence. This is obviously a region where such power corrections are dominant.  Although the sum $(\hat{k}^{(1)}+\hat{k}^{(2)})/\hat{f}$ does not change under different factorization scale choice $\mu$, we take $\mu = E_{H}$ scale choice to remove large logarithms from $\hat{k}^{(2)}$. 

We observe from figure\ \ref{Fig:compare} that the sum $(\hat{k}^{(1)}+\hat{k}^{(2)})/\hat{f}$ can reproduce the NRQCD result quite accurately over a large range of $x_{H}$ already by $Q=30$ GeV. According to eq.\ (\ref{eq:nrqcdmless1}), the ratio of the sum $(\hat{k}^{(1)}+\hat{k}^{(2)})/\hat{f}$ approaches unity as the energy increases at any fixed $x_{H}$.  These plots show 
how accurate the pQCD results are at these energies. To quantify further how closely the pQCD result approaches the NRQCD result, we define $E_\text{down}$ and $E_\text{up}$ as the minimum and the maximum value at which the ratio lies within $0.95 < (\hat{k}^{(1)}+\hat{k}^{(2)})/\hat{f} < 1.05$, respectively.

\bef
\includegraphics[width=3.5in]{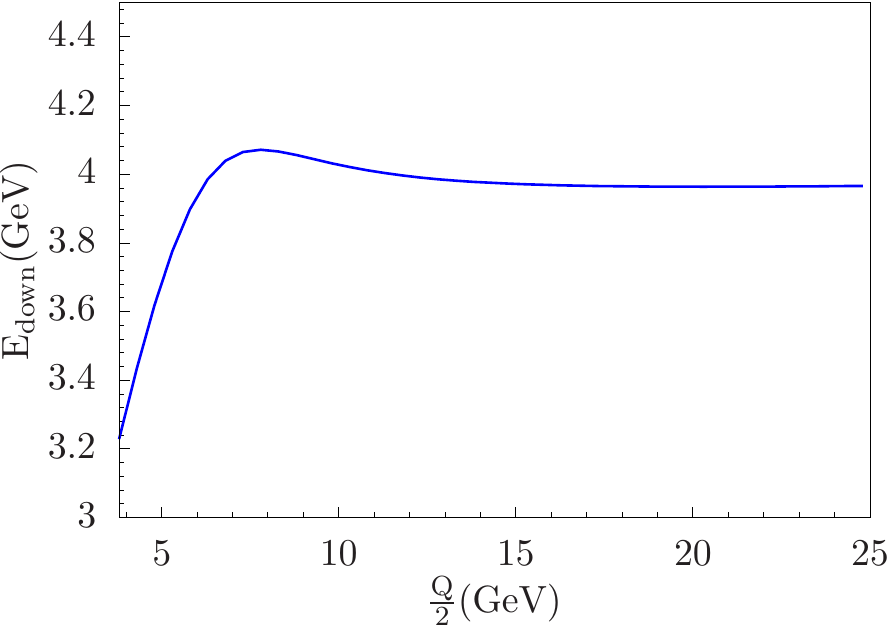} 
\caption{$E_\text{down}$, defined in the text, plotted against $\frac{Q}{2}\approx E_\text{max}$. }
\label{Fig:Edown}
\eef
 
In figure \ref{Fig:Edown}, we plot $E_{\text{down}}$ as a function of $\frac{Q}{2}$. We find it more natural to plot it against $\frac{Q}{2}$, rather than $Q$, since the maximum energy a heavy quarkonium can carry, $E_{\text{max}} = \frac{Q}{2}+\frac{2m_Q^2}{Q}$, is approximately $\frac{Q}{2}$. As shown in figure \ref{Fig:Edown}, the pQCD result approaches NRQCD within 5 percent when $E_{H} \gtrsim 4\, \text{GeV}$ beyond $\frac{Q}{2}$ of about $10$ GeV. Since the minimum energy that a heavy quarkonium can have is $E_{\text{min}} = 2m_Q = 2.8\ \text{GeV}$ in the CM frame, the NLP pQCD result approaches NRQCD rather quickly in this figure of merit.

\bef
\includegraphics[width=3.5in]{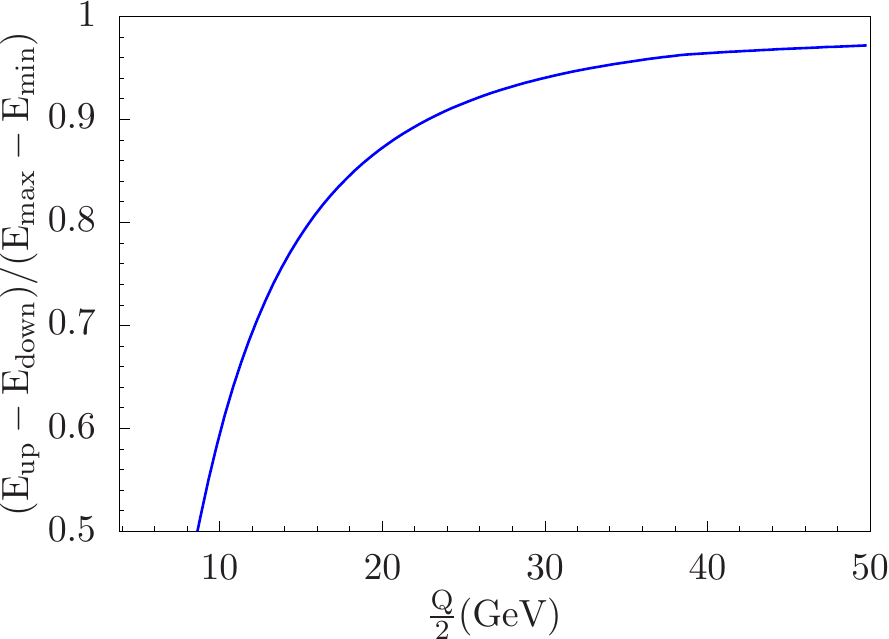} 
\caption{The ratio, $(E_\text{up}-E_\text{down}) / (E_\text{max}-E_\text{min})$, showing the percentage of the energy range that describes NRQCD result within $5$ percent.}
\label{Fig:Emax}
\eef

In figure \ref{Fig:Emax}, we plot $(E_\text{up}-E_\text{down}) / (E_\text{max}-E_\text{min})$ to show the percentage of the energy range that describes the NRQCD result within 5 percent as a function of $\frac{Q}{2}$. When $\frac{Q}{2} \approx 25\ \text{GeV}$, more than $90$ percent of the available range describes NRQCD result within 5 percent.

It is interesting to note that in the case of $p \bar p$ collisions at Tevatron energies, ref.\  \cite{Kang:2014pya} carried out a similar factorized NLP analysis based on an order $\as$ short-distance coefficient, followed by order $\as$ fragmentation.   This is the analog of our $\hat{k}^{(1)}$ term above alone.   In ref.\ \cite{Kang:2014pya}, this term was sufficient to reproduce the $\cc{O}(\alpha_s^2)$ numerical  NRQCD result reasonably well over a large range of $p_T$.  The results here show a similar qualitative agreement from $\hat{k}^{(1)}$, when the factorization scale is chosen as $\mu=E_{H}$.

\subsection{Distribution in $x_{J/\psi}$}
\label{subsec:dist-sH}

Next, we compare the $x_H$ distributions of the various NRQCD channels for $H = J/\psi$, as computed from pQCD factorization through eq.\ (\ref{eq:xHpQCD}). As noted in section \ref{sec:comparison}, there are full calculations only for ${}^3S_1^{[1]}$ and ${}^1S_0^{[8]}$ channels in the literature, so that the remaining pQCD-based curves we will exhibit are in this sense new. Our intention is to compare the shapes, rather than the magnitudes of these contributions to the inclusive cross sections.

Nevertheless, evaluations of the cross sections due to the NRQCD channels require specific values for the long-distance matrix elements. We make the following choices.  For the singlet, we adopt the value from ref.\ \cite{Bodwin:2007fz} of \footnote{Note that there is an additional factor $1/2$ difference in our choice of normalization in the definition of singlet LDME.}
\bea
\langle  \cc{O}^{J/\psi}_{[Q\bar{Q}({}^1S_0^{[8]})]}(\mu_\Lambda)\rangle =\ & 0.22\, ~\text{GeV}^3\,.
\eea
For the octet channels, we take the values suggested as maxima for the treatment of this process in ref.\ \cite{Zhang:2009ym}, except that we also keep a nonzero ${}^3S_1^{[8]}$ matrix element for the purposes of comparison,
\bea
\cc{O}^{J/\psi}_{[Q\bar{Q}({}^3S_1^{[8]})]}(\mu_\Lambda)\rangle\ =\
\langle \cc{O}^{J/\psi}_{[Q\bar{Q}({}^1S_0^{[8]})]}(\mu_\Lambda)\rangle =\ & 2.6\times 10^{-2} ~\text{GeV}^3\, ,\nnu
\langle \cc{O}^{J/\psi}_{[Q\bar{Q}({}^3P_0^{[8]})]}(\mu_\Lambda)\rangle =& \left(\f{m_c^2}{4}\right)2.6\times 10^{-2} ~\text{GeV}^3\, ,
\eea
and
\bea
\langle \cc{O}^{J/\psi}_{[Q\bar{Q}({}^3P_J^{[8]})]}(\mu_\Lambda)\rangle =\ & (2J+1)\langle \cc{O}^{J/\psi}_{[Q\bar{Q}({}^3P_0^{[8]})]}(\mu_\Lambda)\rangle\, . \nnu
\eea
Figure \ref{Fig:CSCO} gives the $x_{J/\psi}$-distributions found from eq.\ (\ref{eq:xHpQCD}) for CM energies of $10.6,\, 30$ and $100$ GeV for these choices of matrix elements.   For 10.6 GeV, we can compare to Belle data \cite{Pakhlov:2009nj}.\footnote{We have transformed their momentum distribution to $x_H$ distribution}  From figure \ref{Fig:CSCO}, we find that the color-singlet distribution has the same general shape as the data, decreasing gently toward zero as $x_{J/\psi}$ approaches unity. 
In contrast, all the non-negligible color-octet curves provide end-point enhancements.  The curves retain these features at the higher energies.  Meaningful comparisons to the data near such endpoints, however, can only be made after organizing large logarithms there \cite{Beneke:1997qw,Fleming:2003gt,Fleming:2006cd,Leibovich:2007vr,Ma:2017xno}.

\bef
\includegraphics[width=6in]{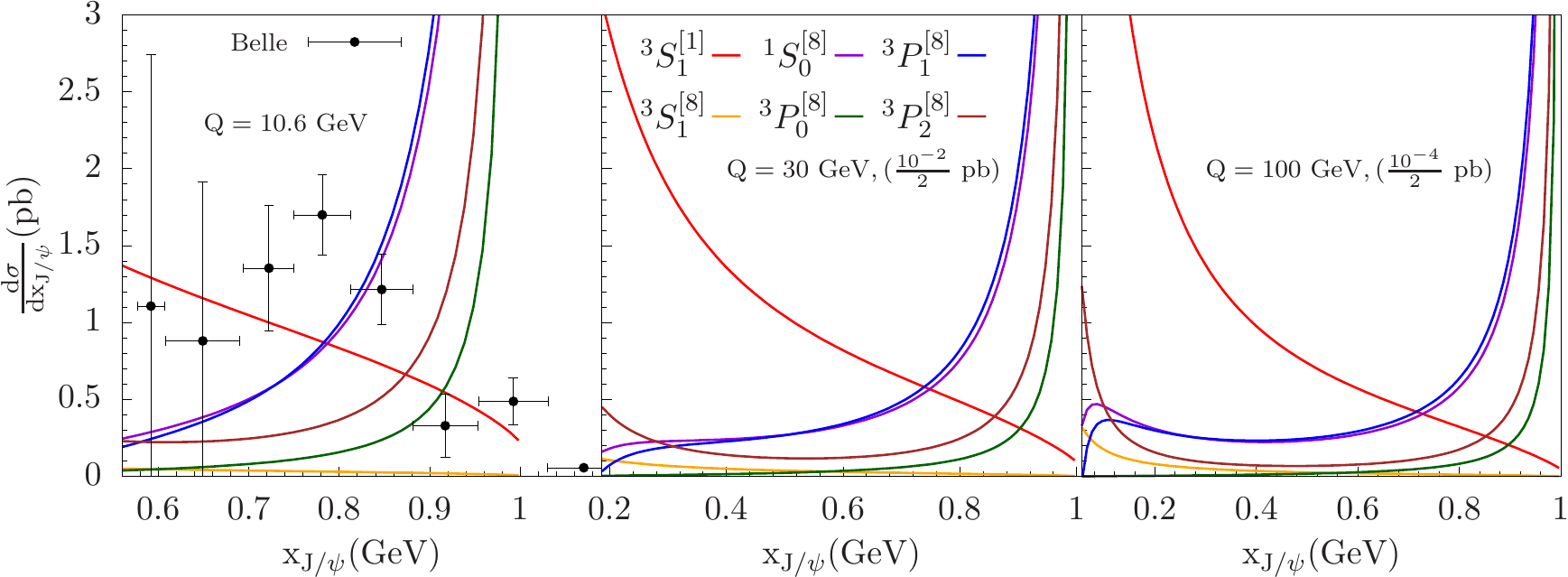} 
\caption{Numerical results for various NRQCD channels at various CM energies $10.6$, $30$, and $100$ GeV with LDME choices described in the text. For $Q=10.6$ GeV, Belle data is shown.}
\label{Fig:CSCO}
\eef

\subsection{Leading logarithms at two-loop order for ${}^3S_1^{[1]}$}
\label{sec:doublelog}
The results of  section\ \ref{sub:approach} suggest that pQCD factorization cross sections approach those of NRQCD rather quickly. The relative ease of computation for pQCD factorization hard parts, where $m_Q $ is taken to be zero, can facilitate the systematic computation of fixed order NRQCD results at high energy.   The pQCD factorization approach, however, provides not only the asymptotic behavior for NRQCD calculations, but also evolution equations \cite{Kang:2014tta} that organize logarithms of $\ln (E_{H}^2/(2m_Q)^2)$ to all orders.

To be specific, the heavy quark pair fragmentation functions satisfy the evolution equations
\bea
\label{eq:QQevol}
\f{\partial}{\partial \ln \mu^2} \mathcal{D}_{[Q\bar{Q}(\kappa)]\to {H}}(z,u,v,\mu;m_Q) =& \sum_{\kappa'}\int_z^1\f{dz'}{z'}\int_0^1 du'\int_0^1 dv'\nnu
&\times \cc{D}_{[Q\bar{Q}(\kappa')]\to {H}}(z',u',v',\mu;m_Q) \nnu
&\times \Gamma_{[Q\bar{Q}(\kappa)]\to [Q\bar{Q}(\kappa')]}(z/z',u,v;u',v',\alpha_s),
\eea
which resums logarithms coming from diagrams in which the pair coherently emits a gluon, bringing it to a new state.   The results for evolution kernels $\Gamma$  are related to the order $\alpha_s$  splitting kernels defined in eq.\ (\ref{eq:partDMS}) by \cite{Kang:2014tta,Fleming:2013qu}\footnote{Singlet to singlet evolution kernels are the well-known Efremov-Radyushkin-Brodsky-Lepage evolution kernels  at $z=1$ \cite{Lepage:1979zb,Efremov:1978rn}.}  
\bea
\Gamma_{[Q\bar{Q}(\kappa)]\to [Q\bar{Q}(\kappa')]} \equiv \left(\f{\alpha_s}{2\pi}\right)P_{\kappa \to \kappa'} \, .
\eea
Solving the pair evolution equations, (\ref{eq:QQevol})  to all orders is nontrivial, and we are not aware of a general formalism for doing so yet in the literature. However, with the tools at hand we can provide explicit results for leading logarithms at two-loop order for the color-singlet channel ${}^3S_1^{[1]}$ to study the importance of resumming the logarithms, $\ln (E_{H}^2/(2m_Q)^2)$. We restrict ourselves to the color-singlet channel case as a test.  Note that at the order we consider for gluon-associated processes, the heavy quark pair can only be created at short distances, and we need only the evolution of the heavy quark pairs among themselves by eq.\ (\ref{eq:QQevol}). The general problem includes mixing between the heavy quark pairs and the single partons  \cite{Kang:2014tta}.

Equation (\ref{eq:QQevol}) has the following two-loop order solution for the fixed heavy pair NRQCD state ${}^3S_1^{[1]}$, with $\beta_0=11N_c/3 - 2N_f/3$,
\bea
\label{eq:twoloopsol}
\cc{D}&_{[Q\bar{Q}(\kappa)] \to [Q\bar{Q}({}^3S_1^{[1]})] \to {J/\psi}}(z,u,v;m_Q,\mu) = \cc{D}_{[Q\bar{Q}(\kappa)] \to [Q\bar{Q}({}^3S_1^{[1]})] \to {J/\psi}}(z,u,v;m_Q,\mu_0)\nnu
&+ \f{\alpha_s(\mu)}{2\pi}\ln \f{\mu^2}{\mu_0^2} \left[1+\f{\beta_0}{2} \f{\alpha_s(\mu)}{4\pi} \ln \f{\mu^2}{\mu_0^2}\right] \nnu
&\hspace{1cm}\times\sum_{\kappa'} P_{\kappa \to \kappa'}\left(\frac{z}{z'},u,v,u',v'\right) \otimes_{z';u',v'} \cc{D}_{[Q\bar{Q}(\kappa')]\to [Q\bar{Q}({}^3S_1^{[1]})] \to {J/\psi}}(z',u',v';m_Q,\mu_0)\nnu
&+ \f{1}{2}\left(\f{\alpha_s(\mu)}{2\pi}\ln \f{\mu^2}{\mu_0^2}\right)^2\sum_{\kappa',\kappa''} P_{\kappa \to \kappa'}\left(\frac{z}{z'},u,v,u',v'\right)\otimes_{z';u',v'}\nnu
&\hspace{0.5cm} P_{\kappa' \to \kappa''}\left(\frac{z'}{z''},u',v',u'',v''\right) \otimes_{z'';u'',v''}  \cc{D}_{[Q\bar{Q}(\kappa'')]\to [Q\bar{Q}({}^3S_1^{[1]})] \to {J/\psi}}(z'',u'',v'';m_Q,\mu_0)\, ,
\eea
which can be easily checked perturbatively \cite{Bodwin:2014bpa,Jia:2008ep}. As appropriate to our factorization, we set our input scale $\mu_0 = 2 m_Q$ and the hard scale $\mu = E_{H}$. These choices remove large logarithms from the short-distance coefficients and keep them in the fragmentation functions. Their ratio specifies the logarithms we would like to resum, $\ln (E_{H}^2/(2m_Q)^2)$.  The fragmentation functions at the input scale $\mu_0$ are found as in eq.\ (\ref{inputFF_dec}), where we now specify the NRQCD state $\nu$ there to be ${}^3S_1^{[1]}$,
\bea
\cc{D}_{[Q\bar{Q}(\kappa)]\to  [Q\bar{Q}({}^3S_1^{[1]})]]\to H }(z,u,v;m_Q,\mu_0)=&\bigg(\hat{d}^{(0)}_{[Q\bar{Q}(\kappa)]\to [Q\bar{Q}({}^3S_1^{[1]})]]} (z,u,v; m_Q, \mu_0,\mu_\Lambda) \nnu
&\hspace{-55mm}+ \left(\f{\as}{\pi}\bigg)\hat{d}^{(1),\overline{\text{MS}}}_{[Q\bar{Q}(\kappa)]\to [Q\bar{Q}({}^3S_1^{[1]})]]}(z,u,v;m_Q, \mu_0,\mu_\Lambda) + \mathcal{O}(\alpha_s^2)\right)\times\f{\langle \cc{O}^{H}_{[Q\bar{Q}(\nu)]}(\mu_\Lambda)\rangle}{m_Q^{2L+1}}\,.
\eea
As we only concern ourselves with the leading logarithms, unsuppressed by further powers of $\alpha_s$, we use the tree level matching $\hat{d}^{(0)}$. Since the only nonzero $\hat{d}^{(0)}_{[Q\bar{Q}(\kappa)]\to [Q\bar{Q}({}^3S_1^{[1]})]]}$ is $\kappa = v[1]$, the final pQCD state in the kernel must always be $v[1]$. 

To get the $x_H $ distribution, we must do a further convolution $\otimes_{z;u,v}$ of the two-loop solution in eq.\ (\ref{eq:twoloopsol}) with short-distance coefficients computed in section \ref{sec:lo-nlo}. Again, because we do not want further suppression in $\alpha_s$, we only need to convolve with the leading-order short distance coefficients found in eq.\ (\ref{eq:sig-1-v[8]}) and (\ref{eq:sig-1-a[8]}),
\bea
\f{d\hat{\sigma}^{(1)}_{e^+e^- \to Q\bar{Q}(v[8])+g}}{dx}\left(x=\frac{x_H}{z},u,v\right) \hspace{0.5cm} \text{and} \hspace{0.5cm} \f{d\hat{\sigma}^{(1)}_{e^+e^- \to Q\bar{Q}(a[8])+g}}{dx}\left(x=\frac{x_H}{z},u,v\right)\,,
\eea
and thus $\kappa$ in eq.\eqref{eq:twoloopsol} must be either $v[8]$ or $a[8]$. From now on, we suppress the arguments of the short-distance coefficients, splitting functions, and fragmentation functions for simplicity.

Calculating the single logarithms from the perturbative solution gives an expected result, in agreement
with eq.\ (\ref{eq:k13S1}),
\bea
\left(\f{\alpha_s}{2\pi} \ln \f{x_{H}^2}{4r}\right)\,&\f{d\hat{\sigma}^{(1)}_{e^+e^- \to Q\bar{Q}(v[8])+g}}{dx} \otimes_{z;u,v} P_{v[8] \to v[1]} \otimes_{z';u',v'} \cc{D}^{(0)}_{[Q\bar{Q}(v[1])]\to [Q\bar{Q}({}^3S_1^{[1]})] \to {H}} \nnu
=& \left(\f{\alpha_s}{2\pi} \ln \f{x_{H}^2}{4r}\right)\,\f{d\hat{\sigma}^{(1)}_{e^+e^- \to Q\bar{Q}(a[8])+g}}{dx} \otimes_{z;u,v} P_{a[8] \to v[1]} \otimes_{z';u',v'} \cc{D}^{(0)}_{[Q\bar{Q}(v[1])]\to [Q\bar{Q}({}^3S_1^{[1]})] \to {H}} \nnu
=&\sigma_0  \f{\alpha_s^2e_Q^2}{Q^2}\f{128}{9}\f{(1-x_{H})}{x_{H}(2-x_{H})^2} \ln\left(\f{x_{H}^2}{4r}\right) \f{\langle \cc{O}^{H}_{[Q\bar{Q}({}^3S_1^{[1]})]} \rangle}{m_Q} \,,
\eea
where for convenience we have rewritten $\ln (E_{H}^2/(2m_Q)^2)$ as $\ln (x_{H}^2/(4r))$.
At the next order, there are double log terms associated with the running of the coupling in these terms, found simply by multiplying additional factors
\bea
\f{\beta_0}{2} \f{\alpha_s}{4\pi} \ln \f{x_{H}^2}{4r}
\eea
as can be seen from eq.\ (\ref{eq:twoloopsol}).

Less trivial calculations involve the convolution of two evolution kernels, and we simply report the results of our computation
\bea
\f{1}{2}\left(\f{\alpha_s}{2\pi}\ln \f{x_{H}^2}{4r}\right)^2&\sum_{\kappa'} \f{d\hat{\sigma}^{(1)}_{e^+e^- \to Q\bar{Q}(v[8])+g}}{dx} \otimes_{z;u,v} P_{v[8] \to \kappa'} \otimes_{z';u',v'} P_{\kappa' \to v[1]} \otimes_{z'';u'',v''} \cc{D}^{(0)}_{[Q\bar{Q}(v[1])]\to [Q\bar{Q}({}^3S_1^{[1]})] \to {H}} \nnu
=\ &\f{1}{2}\left(\f{\alpha_s}{2\pi}\ln \f{x_{H}^2}{4r}\right)^2 \sigma_0 \f{\alpha_se_Q^2}{Q^2 } \f{16\pi}{27} \frac{\langle \cc{O}^{H}_{[Q\bar{Q}({}^3S_1^{[1]})]} \rangle}{m_Q} \f{1}{x_{H}(2-x_{H})^2}\nnu
&\times \bigg[\left(9x_{H}^4+18x_{H}^3-54x_{H}^2+152x_{H}-52\right)\ln x_{H} + 32\left(x_{H}^2+x_{H}-2\right)\ln(1-x_{H}) \nnu
&\hspace{-2.5cm}-\left(9x_{H}^4+18x_{H}^3-22x_{H}^2-24x_{H}+124\right) \ln(2-x_{H})+6(1-x_{H})\left(-21x_{H}^2+36x_{H}+28+8\ln 2\right)\bigg]\,,\\
\f{1}{2}\left(\f{\alpha_s}{2\pi}\ln \f{x_{H}^2}{4r}\right)^2&\sum_{\kappa'} \f{d\hat{\sigma}^{(1)}_{e^+e^- \to Q\bar{Q}(a[8])+g}}{dx} \otimes_{z;u,v} P_{a[8] \to \kappa'} \otimes_{z';u',v'} P_{\kappa' \to v[1]} \otimes_{z'';u'',v''} \cc{D}^{(0)}_{[Q\bar{Q}(v[1])]\to [Q\bar{Q}({}^3S_1^{[1]})] \to {H}} \nnu
=\ &\f{1}{2}\left(\f{\alpha_s}{2\pi}\ln \f{x_{H}^2}{4r}\right)^2 \sigma_0 \f{\alpha_se_Q^2}{Q^2} \f{16\pi}{27} \frac{\langle \cc{O}^{H}_{[Q\bar{Q}({}^3S_1^{[1]})]} \rangle}{m_Q}\f{1}{x_{H}(2-x_{H})^2}\nnu
&\times \bigg[-\left(9x_{H}^4-134x_{H}^2+276x_{H}-196\right)\ln x_{H} + 32\left(x_{H}^2+x_{H}-2\right)\ln(1-x_{H}) \nnu
&\hspace{-2cm}+\left(9x_{H}^4-166x_{H}^2+132x_{H}-52\right) \ln(2-x_{H})-2(1-x_{H})\left(9x_{H}^2+6x_{H}-120-8\ln 2\right)\bigg]\, .
\eea
Physically, these contributions describe initial octet states, $v[8]$ or $a[8]$, evolving through the allowed states $\kappa'$, which themselves evolve to $v[1]$, then to the color-singlet NRQCD state ${}^3S_1^{[1]}$, and finally to hadron ${H}$. 
We can also vary the factorization scale from $E_{H}/2$ to $2E_{H}$.   We choose the renormalization and factorization scales to be equal.

In figure \ref{Fig:ln2}, we show numerically the impact of including the two-loop terms in the leading logarithmic series.   For ease of comparison, we will continue to report our results as in figure \ref{Fig:compare}, as ratios with the full NRQCD results given in eq.\ (\ref{eq:nrqcdpred}). As noted before, the divergence as $x_{H} \to 4m_Q/Q$ is due to $\hat{f}$ vanishing in the lower limit, where the pQCD result does not apply. Since $E_{H}$ increases as $Q$ increases at a fixed $x_{H}$, the logarithms $\ln (E_{H}^2/(2m_Q)^2)$ become larger at higher $Q$ for a fixed $x_{H}$. At Belle's energy of $Q = 10.6$ \text{GeV}, $E_{H}$ and $2m_Q$ do not create a strong hierarchy and the resummation of such logarithms is not so important. However, as can be seen from figure \ref{Fig:ln2}, the leading logarithms at two-loop order modify the results up to $\sim 30\%$ already at $Q=30$ GeV. Such a large correction implies that solving the evolution equation to resum the logarithms $\ln (E_{H}^2/(2m_Q)^2)$ is crucial for a reliable prediction at higher energies than at Belle.

\bef
\includegraphics[width=6in]{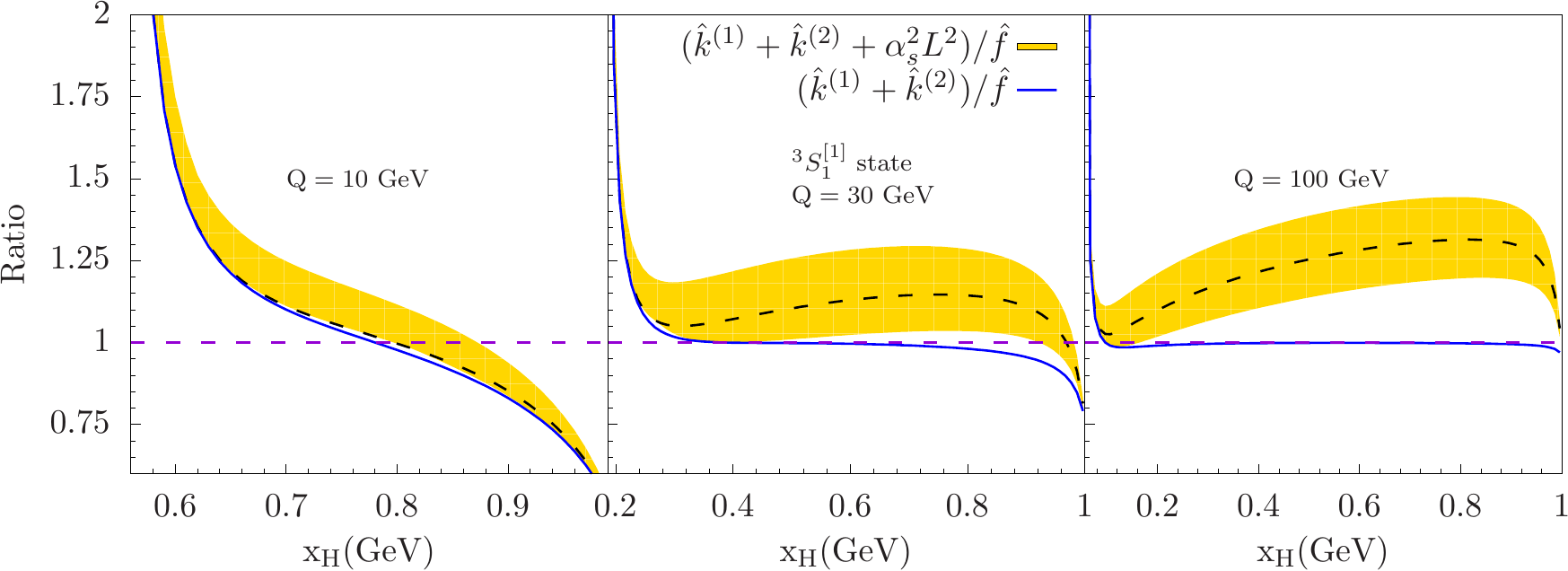} 
\caption{ The effect of including the leading logarithm from evolution equation (\ref{eq:QQevol}) on the ratios of the full pQCD and NRQCD results, $(\hat{k}^{(1)}+\hat{k}^{(2)})/\hat{f}$ for the ${}^3S_1^{[1]}$ channel at CM energies $10$, $30$, and $100$ GeV.}
\label{Fig:ln2}
\eef

\section{Conclusions}
In this paper, we presented the first NLO calculation of short-distance coefficients in the context of NLP perturbative QCD factorization for heavy quarkonia, extending the quark pair fragmentation formalism developed in ref. \cite{Kang:2014tta}. We calculated short-distance coefficients as closed expressions for the physically-relevant $x_H$ distribution of heavy quarkonia in $\rm e^+e^-$ annihilation.  We showed that it is useful to include evanescent operators, which are absent in the four-dimensional theory, when using dimensional regularization.  At NLO, the contribution from the evanescent operators organizes finite pieces that are associated with long-distance dynamics.    This organization is important to clarify comparisons with fixed-order NRQCD calculations.

Combining our calculation of NLO short-distance coefficients with NRQCD factorization of the heavy quark pair fragmentation functions \cite{Ma:2013yla,Ma:2014eja}, we derived the high energy behavior of the $x_H$ distributions in fixed-order NRQCD for various channels. We presented numerical results of these channels at various CM energies, which exhibit end-point enhancements for the octet channels and a large contribution from the singlet channel away from the tail. Using the singlet channel, we numerically illustrated that fixed-order NRQCD results rapidly approach their high energy behavior and demonstrated the importance of resumming logarithms of $\ln (E_H^2/(2m_Q)^2)$ by studying the numerical impact of including the leading logarithms at two-loop order. 

We also showed explicit analytical agreement between our derived high energy behavior and the fixed-order NRQCD calculations in the literature for the ${}^3S_1^{[1]}$  \cite{Cho:1996cg,Keung:1980ev,Yuan:1996ep} and ${}^1S_0^{[8]}$ \cite{Sun:2018yam} channels, illustrating the need for including the evanescent operators identified. The formalism developed in our work provides the groundwork for explicit NLO and higher order calculations of short-distance coefficients for many other factorized cross sections at next-to-leading power.   We anticipated applications to angular distributions in $\rm e^+e^-$ annihilation, and to many single-particle cross sections for heavy quarkonia in hadron-hadron scattering.

\acknowledgments
We thank Z.-B. Kang, Y.-Q. Ma, J.-W. Qiu, W. Vogelsang, and H. Zhang for helpful discussions. We would like to thank Y.-Q. Ma for also providing his code to calculate input fragmentation functions using NRQCD. This research was supported by the National Science Foundation under Grants PHY-1316617, 1620628 and 1915093.

\appendix
\section{$D-4$ dimensional components in NLO cross sections}
\label{appendixPS}
Our NLO process $e^+e^- \to Q\bar{Q}(p_1) + g(p_2) + g(p_3)$ needs to take $D-4$ dimensional momentum components into account correctly. Since the heavy quark pair is observed, $p_1$ is a $4$ dimensional momentum, whereas $p_2$ and $p_3$ are not. That is, 
\bea
p_1 =& p_{1,4}\,,\\
p_2 =& p_{2,4} + \hat{p}_2\,, \\
p_3 =& p_{3,4} - \hat{p}_2\,,
\eea
where $p_{i,4}$ and $\hat{p}_i$ are $4$-dimensional and $D-4$ dimensional components of the momentum $p_i$, respectively. 

The phase space factor from the eq.\ (\ref{eq:phasespace}) can then be simplified to
\bea
\label{eq:apdixPS}
d\Pi_3 = &\frac{d^{D-1}p_1}{(2\pi)^{D-1}2E_1}\,\frac{d^{D-1}p_2}{(2\pi)^{D-1}2E_2}\,\frac{d^{D-1}p_3}{(2\pi)^{D-1}2E_3}(2\pi)^D \delta^D(q-p_1-p_2-p_3)\nnu
=& \frac{d^{D-1}p_1}{(2\pi)^{D-1}2E_1}\,\frac{d^{D-1}p_2}{(2\pi)^{D-1}2E_2} 2\pi \delta\left(Q^2(1-x_1-x_2 + \frac{1}{2}x_1 x_2(1-\cos \theta_{1}))\right)\,,
\eea
where $x_i = 2E_i/Q$ and $\theta_{1}$ is the angle in the $D-1$ plane between the spatial components of momentum $p_1$ and $p_2$, $\vec{p}_1$ and $\vec{p}_2$. The $D-1$ plane consists of `$x$-$y$-$z$' axes and axes of the remaining $D-4$ hyperplane. $\hat{p}_2$ is in the $D-4$ hyperplane and can be projected out from the $D-1$ plane by the following procedure. We first choose $\vec{p}_1$ to point along the $z$-axis of the $D-1$ plane. Then the $D-2$ dimensional transverse part of $\vec{p}_2$ is $|\vec{p}_2| \sin \theta_{1}$. We then define $\theta_2$ as the angle such transverse vector makes with respect to the $y$-axis of the $D-2$ plane. Then $D-3$ dimensional transverse piece can be projected out by $|\vec{p}_2| \sin \theta_1 \sin \theta_2$. Finally, we define $\theta_3$ to represent the angle that the $D-3$ dimensional transverse piece makes with respect to the $x$-axis, giving us
\bea
\hat{p}_2^2 = |\vec{p}_2|^2 \sin^2 \theta_1 \sin^2 \theta_2 \sin^2 \theta_3 =  Q^2\,(1-x_1)\,y\,(1-y)\,\sin^2 \theta_2\sin^2 \theta_3\,,
\eea
where the last equality used the $\delta$ function in eq.\ (\ref{eq:apdixPS}) and $y = (1-x_2)/x_1$.

Working out the above phase space factor in eq.\ (\ref{eq:apdixPS}) in terms of these angles, we arrive at
\bea
\label{eq:finalPS}
d\Pi_3 =& \f{Q^2}{16(2\pi)^3}\left(\f{4\pi}{Q^2}\right)^{2\epsilon}\f{1}{\Gamma(2-2\epsilon)}y^{-\epsilon}\,(1-y)^{-\epsilon}\,(1-x_1)^{-\epsilon}\,x_1^{1-2\epsilon}\,dx_1\,dy \nnu
\times&\f{1}{\pi}\f{\Gamma\left(\f{D-2}{2}\right)}{\Gamma\left(\f{D-4}{2}\right)}\int d\theta_2\,\int d\theta_3\,\sin^{D-4}\theta_2\,\sin^{D-5}\theta_3\,\int d\hat{p}_2^2\,\delta\left(\hat{p}_2^2 - Q^2\,(1-x_1)\,y\,(1-y)\,\sin^2 \theta_2\sin^2 \theta_3\right)\,.
\eea
The second line integrates to unity for $\hat{p}_2^2$ independent terms, but integrates to
\bea
Q^2(1-x_1)y(1-y)\left(\f{-2\epsilon}{2-2\epsilon}\right)
\eea
for the integrand with $\hat{p}_2^2$. Therefore, we can replace $\hat{p}_2^2$ as in eq.\ (\ref{eq:p2replace}) and use the first line of eq.\ (\ref{eq:finalPS}) as the phase space factor as in eq.\ (\ref{eq:Pi3}).

\section{Matching coefficients of input heavy quark pair fragmentation functions in BMHV scheme}
\label{appendixhatd}
In this appendix, we summarize all the matching coefficients of input heavy quark pair fragmentation functions used in this paper with BMHV $\gamma_5$ and $\overline{\text{MS}}$ subtraction scheme. The calculation of the same coefficients with Kreimer $\gamma_5$ and $\overline{\text{MS}}$ subtraction scheme can be found in the appendices of \cite{Ma:2013yla,Ma:2014eja}. The additional terms in BMHV $\gamma_5$ scheme are indicated by the square brackets below.
\bea
\hat{d}^{(0)}_{[Q\bar{Q}(v[1])]\to [Q\bar{Q}({}^3S_1^{[1]})]} =& \frac{1}{24}\delta(u-\frac{1}{2})\delta(v-\frac{1}{2})\delta(1-z)\,,\\
\hat{d}^{(0)}_{[Q\bar{Q}(v[8])]\to [Q\bar{Q}({}^3S_1^{[8]})]} =& \frac{1}{192}\delta(u-\frac{1}{2})\delta(v-\frac{1}{2})\delta(1-z)\,,\\
\hat{d}^{(0)}_{[Q\bar{Q}(a[8])]\to [Q\bar{Q}({}^1S_0^{[8]})]}=& \frac{1}{64}\delta(u-\frac{1}{2})\delta(v-\frac{1}{2})\delta(1-z)\,,\\
\hat{d}^{(0)}_{[Q\bar{Q}(v[8])]\to [Q\bar{Q}({}^3P_0^{[8]})]} =& \frac{1}{768}\delta'(u-\frac{1}{2})\delta'(v-\frac{1}{2})\delta(1-z)\,,\\
\hat{d}^{(0)}_{[Q\bar{Q}(a[8])]\to [Q\bar{Q}({}^3P_1^{[8]})]} =& \frac{1}{96}\delta(u-\frac{1}{2})\delta(v-\frac{1}{2})\delta(1-z)\,,\\
\hat{d}^{(0)}_{[Q\bar{Q}(v[8])]\to [Q\bar{Q}({}^3P_2^{[8]})]} =& \frac{1}{1920}\delta'(u-\frac{1}{2})\delta'(v-\frac{1}{2})\delta(1-z)\,,\\
\hat{d}^{(1),\overline{\text{MS}}}_{[Q\bar{Q}(v[8])]\to [Q\bar{Q}({}^3S_1^{[1]})], z\neq 1} =& \frac{C_F}{24}\frac{1}{N_c^2-1}\Delta_{-}^{[1]}\frac{z}{(1-z)}\left(\ln\frac{\mu_0^2}{4m_Q^2(1-z)^2}+2z^2-4z+1\right)\,,\\
\hat{d}^{(1),\overline{\text{MS}}}_{[Q\bar{Q}(v[8])]\to [Q\bar{Q}({}^3S_1^{[8]})], z\neq 1} =& \frac{C_F}{24}\frac{1}{(N_c^2-1)^2}\Delta_{-}^{[8]}\frac{z}{(1-z)}\left(\ln\frac{\mu_0^2}{4m_Q^2(1-z)^2}+2z^2-4z+1\right)\,,\\
\hat{d}^{(1),\overline{\text{MS}}}_{[Q\bar{Q}(v[8])]\to [Q\bar{Q}({}^1S_0^{[8]})], z\neq 1} =& \frac{C_F}{8}\frac{1}{(N_c^2-1)^2}\Delta_+^{[8]}z(1-z)\left(\ln\frac{\mu_0^2}{4m_Q^2(1-z)^2}-3 +[3]\right)\,,\\
\hat{d}^{(1),\overline{\text{MS}}}_{[Q\bar{Q}(v[8])]\to [Q\bar{Q}({}^3P_0^{[8]})], z\neq 1} =& \frac{C_F}{24}\frac{1}{(N_c^2-1)^2}z\bigg\{\Delta_{+}^{[8]}(1-z)\left(\ln\frac{\mu_0^2}{4m_Q^2(1-z)^2}+\frac{5z^2-3}{3(1-z)^2}\right) \nnu
&\hspace{-1cm}+\Delta_{+}^{[8]'}\left(\ln\frac{\mu_0^2}{4m_Q^2(1-z)^2}-\frac{1}{1-z}\right) +\frac{\Delta_{+}^{[8]''}}{1-z}\left(\ln\frac{\mu_0^2}{4m_Q^2(1-z)^2}-1\right)\bigg\} \\
\hat{d}^{(1),\overline{\text{MS}}}_{[Q\bar{Q}(v[8])]\to [Q\bar{Q}({}^3P_1^{[8]})], z\neq 1} =& \frac{C_F}{12}\frac{1}{(N_c^2-1)^2}z\bigg\{\Delta_{+}^{[8]}(1-z)\left(\ln\frac{\mu_0^2}{4m_Q^2(1-z)^2}+\frac{3-z^2}{6(1-z)^2}\right)\nnu
&+\frac{\Delta_{+}^{[8]'}}{2}(\frac{3}{2}-z)+\frac{\Delta_{+}^{[8]''}}{2}(1-z)\bigg\}\,,\\
\hat{d}^{(1),\overline{\text{MS}}}_{[Q\bar{Q}(v[8])]\to [Q\bar{Q}({}^3P_2^{[8]})], z\neq 1} =&  \frac{C_F}{60}\frac{1}{(N_c^2-1)^2}z\bigg\{\Delta_{+}^{[8]}(1-z)\left(\ln\frac{\mu_0^2}{4m_Q^2(1-z)^2}+\frac{7z^2+3}{6(1-z)^2}\right)\nnu
&+\Delta_{+}^{[8]'}\left(\ln\frac{\mu_0^2}{4m_Q^2(1-z)^2}+\frac{1}{4(1-z)}(6z^2-3z-7)\right)\nnu
&+\frac{\Delta_{+}^{[8]''}}{1-z}\left(\ln\frac{\mu_0^2}{4m_Q^2(1-z)^2}+\frac{1}{2}(3z^2-6z+1)\right)\bigg\}\,,\\
\hat{d}^{(1),\overline{\text{MS}}}_{[Q\bar{Q}(a[8])]\to [Q\bar{Q}({}^3S_1^{[1]})], z\neq 1}  =& \frac{C_F}{24}\frac{1}{N_c^2-1}\Delta_{+}^{[1]}z(1-z)\left(\ln\frac{\mu_0^2}{4m_Q^2(1-z)^2}-1+[3]\right) \,,\\
\hat{d}^{(1),\overline{\text{MS}}}_{[Q\bar{Q}(a[8])]\to [Q\bar{Q}({}^3S_1^{[8]})], z\neq 1}  =& \frac{C_F}{24}\frac{1}{(N_c^2-1)^2}\Delta_{+}^{[8]}z(1-z)\left(\ln\frac{\mu_0^2}{4m_Q^2(1-z)^2}-1+[3]\right) \,,\\
\hat{d}^{(1),\overline{\text{MS}}}_{[Q\bar{Q}(a[8])]\to [Q\bar{Q}({}^1S_0^{[8]})], z\neq 1} =& \frac{C_F}{8}\frac{1}{(N_c^2-1)^2}\Delta_-^{[8]}\frac{z}{1-z}\left(\ln\frac{\mu_0^2}{4m_Q^2(1-z)^2}-1\right)\,,\\
\hat{d}^{(1),\overline{\text{MS}}}_{[Q\bar{Q}(a[8])]\to [Q\bar{Q}({}^3P_0^{[8]})], z\neq 1} =& \frac{C_F}{24}\frac{1}{(N_c^2-1)^2}z(1-z)\left(\Delta_{-}^{[8]}+\Delta_{-}^{[8]'}+\Delta_{-}^{[8]''}\right)\left(\ln\frac{\mu_0^2}{4m_Q^2(1-z)^2}-3+[3]\right)\,,\\
\hat{d}^{(1),\overline{\text{MS}}}_{[Q\bar{Q}(a[8])]\to [Q\bar{Q}({}^3P_1^{[8]})], z\neq 1} =& \frac{C_F}{12}\frac{1}{(N_c^2-1)^2}z\bigg\{\frac{\Delta_{-}^{[8]}}{1-z}\left(\ln\frac{\mu_0^2}{4m_Q^2(1-z)^2}-\frac{1}{2}(5z^2-12z+9)+\left[4(1-z)^2\right]\right)\nnu
&+\frac{3}{4}\Delta_{-}^{[8]'}(1-z)+\frac{1}{2}\Delta_{-}^{[8]''}(1-z)\bigg\}\,,\\
\hat{d}^{(1),\overline{\text{MS}}}_{[Q\bar{Q}(a[8])]\to [Q\bar{Q}({}^3P_2^{[8]})], z\neq 1} =&  \frac{C_F}{60}\frac{1}{(N_c^2-1)^2}z(1-z)\bigg\{\left(\Delta_{-}^{[8]}+\Delta_{-}^{[8]''}\right)\left(\ln\frac{\mu_0^2}{4m_Q^2(1-z)^2}-\frac{3}{2}+[3]\right)\nnu
&+\Delta_{-}^{[8]'}\left(\ln\frac{\mu_0^2}{4m_Q^2(1-z)^2}-\frac{15}{4}+[3]\right)\bigg\}\,.
\eea

\section{Results of pQCD short-distance coefficients convolved with input heavy quark fragmentation functions}
\label{appendixkterms}
In this appendix, we give the full results of high energy behavior of NRQCD for various channels derived by convolution of perturbative QCD short-distance coefficients with the input heavy quark fragmentation functions. The superscripts $(i)$ in $\hat{k}^{(i)}$, an explicit definition of which is given in eq.\ (\ref{eq:xHpQCD}), indicate the order of $\alpha_s$ for the short-distance coefficients.
\bea
\hat{k}^{(1)}_{{}^3S_1^{[1]}}(x_H;\mu_0,\mu_\Lambda) =& \frac{256}{9} \f{(1-x_H)}{x_H(2-x_H)^2}\left(  \ln \left( \f{\mu_0^2}{4m_Q^2(1-x_H)^2} \right) +x_H^2-2x_H+\frac{3}{2}\right)\,,\\
\hat{k}^{(2)}_{{}^3S_1^{[1]}}(x_H;\mu_0,\mu_\Lambda) =& -\frac{256}{9} \f{(1-x_H)}{x_H(2-x_H)^2}\left(  \ln \left( \f{\mu_0^2}{E_H^24(1-x_H) } \right) -\frac{x_H^2+2x_H-2}{4(1-x_H)}\right)\,,\\
\hat{k}^{(1)}_{{}^1S_0^{[8]}}(x_H;\mu_0,\mu_\Lambda) =&\f{2}{x_H(1-x_H)(2-x_H)^2}\bigg(12\left((1-x_H)^4+1\right)\ln \left( \f{\mu_0^2}{4m_Q^2(1-x_H)^2} \right) -12\bigg)\,,\\
\hat{k}^{(2)}_{{}^1S_0^{[8]}}(x_H;\mu_0,\mu_\Lambda) =&- \f{2}{x_H(1-x_H)(2-x_H)^2}\bigg(  12\left((1-x_H)^4+1\right)\ln \left( \f{\mu_0^2}{E_H^24(1-x_H) } \right) \nnu
&\hspace{3cm}+ 23x_H^4-78x_H^3+102x_H^2-48x_H+12\bigg)\,,\\
\hat{k}^{(2)}_{{}^3S_1^{[8]}}(x_H;\mu_0,\mu_\Lambda) =&\f{5}{16} k^{(1)}_{{}^3S_1^{[1]}}(x_H;\mu_0,\mu_\Lambda)\,,\\
\hat{k}^{(1)}_{{}^3P_0^{[8]}}(x_H;\mu_0,\mu_\Lambda) =&\f{8x_H}{(1-x_H)(2-x_H)^4}\bigg(\frac{1}{3}\left(5x_H^4-30x_H^3+65x_H^2-60x_H+17\right)\nnu
&+(x_H^4-6x_H^3+16x_H^2-20x_H+10)\ln \left( \f{\mu_0^2}{4m_Q^2(1-x_H)^2} \right) \bigg)\,,\\
\hat{k}^{(2)}_{{}^3P_0^{[8]}}(x_H;\mu_0,\mu_\Lambda) =&-\f{8x_H}{(1-x_H)(2-x_H)^4}\bigg(\frac{1}{12}\left(11x_H^4-36x_H^3+72x_H^2-72x_H+36\right)\nnu
&+(x_H^4-6x_H^3+16x_H^2-20x_H+10)\ln \left( \f{\mu_0^2}{E_H^24(1-x_H)} \right) \bigg)\,,\\
\hat{k}^{(1)}_{{}^3P_1^{[8]}}(x_H;\mu_0,\mu_\Lambda) =&\f{4}{3}\f{1}{x_H(1-x_H)(2-x_H)^2}\bigg(12((1-x_H)^4+1)\ln \left( \f{\mu_0^2}{4m_Q^2(1-x_H)^2} \right)\nnu
& -2\f{\left(4x_H^6-33x_H^5+100x_H^4-159x_H^3+154x_H^2-108x_H+48\right)}{(2-x_H)^2}\bigg)\,,\\
\hat{k}^{(2)}_{{}^3P_1^{[8]}}(x_H;\mu_0,\mu_\Lambda) =&- \f{4}{3}\f{1}{x_H(1-x_H)(2-x_H)^2}\bigg(  12\left((1-x_H)^4+1\right)\ln \left( \f{\mu_0^2}{E_H^24(1-x_H) } \right) \nnu
&\hspace{3cm}+ 23x_H^4-78x_H^3+102x_H^2-48x_H+12\bigg)\,,\\
\hat{k}^{(1)}_{{}^3P_2^{[8]}}(x_H;\mu_0,\mu_\Lambda) =&\f{16}{5}\f{x_H}{(1-x_H)(2-x_H)^4}\bigg((x_H^4-6x_H^3+16x_H^2-20x_H+10)\ln \left( \f{\mu_0^2}{4m_Q^2(1-x_H)^2} \right)\nnu
& -\frac{1}{6x_H^2}\left(2x_H^6-3x_H^5-100x_H^4+453x_H^3-814x_H^2+684x_H-216\right)\bigg)\,,\\
\hat{k}^{(2)}_{{}^3P_2^{[8]}}(x_H;\mu_0,\mu_\Lambda) =&-\f{16}{5}\f{x_H}{(1-x_H)(2-x_H)^4}\bigg(\frac{1}{12}\left(11x_H^4-36x_H^3+72x_H^2-72x_H+36\right)\nnu
&+(x_H^4-6x_H^3+16x_H^2-20x_H+10)\ln \left( \f{\mu_0^2}{E_H^24(1-x_H)} \right)\bigg)\,.
\eea
\bibliographystyle{JHEP}
\bibliography{bibliography}
\end{document}